\documentclass[sn-mathphys,Numbered]{sn-jnl}

\usepackage{graphicx}%
\usepackage{multirow}%
\usepackage{amsmath,amssymb,amsfonts}%
\usepackage{amsthm}%
\usepackage{mathrsfs}%
\usepackage[title]{appendix}%
\usepackage{xcolor}%
\usepackage{textcomp}%
\usepackage{manyfoot}%
\usepackage{booktabs}%
\usepackage{algorithm}%
\usepackage{algorithmicx}%
\usepackage{algpseudocode}%
\usepackage{listings}%
\usepackage{amsthm}
\usepackage{amsmath} 
\usepackage{amsfonts}
\usepackage{amssymb}
\usepackage{booktabs}

\def\bP{\mathbb{P}}
\def\bE{\mathbb{E}}

\theoremstyle{thmstyleone}%
\newtheorem{theorem}{Theorem}
\theoremstyle{thmstyletwo}%

\theoremstyle{thmstylethree}%
\newtheorem{lemma}{Lemma}

\usepackage{bbm} 
\usepackage[shortlabels]{enumitem}
\begin{document}

\title{Linearly-scalable and entropy-optimal learning of nonstationary and nonlinear manifolds}


\author*[1]{\fnm{Illia} \sur{Horenko}}\email{horenko@rptu.de}


\affil*[1]{\orgdiv{Chair for Mathematics of AI, Faculty of Mathematics}, \orgname{RPTU Kaiserslautern-Landau}, \orgaddress{\street{Gottlieb-Daimler-Str. 48}, \city{Kaiserslautern}, \postcode{67663}, \country{Germany}}\vspace{-0.5cm}}

\abstract{
Unsupervised extraction of relevant low-dimensional manifolds from high-dimensional data is in core of many data analysis problems. Common linear methods like the Principal Component Analysis (PCA) and related linear approaches  scale linearly with the data statistics size $T$ - but frequently fail to extract the nonlinear or changing in time (nonstationary) low-dimensional manifolds.  Common nonlinear methods 
scale as $\mathcal{O}(T^2)$, or  $\mathcal{O}(T\textrm{log}(T))$ in the best case, frequently fail extracting nonstationary manifolds,   and can distort results by generating method-induced artefacts (e.g., reveal clusters that are actually not present in the data, or do not reveal clusters when they are present, some examples are provided). 
Here we propose  an Entropy-Optimal Manifold Clustering (EOMC) as a metricised and entropy-regularized extension of PCA clustering - and show that it mitigates the problems of the existing tools even in very nonstationary and nonlinear situations, while pertaining the favourable  $\mathcal{O}(T)$ iteration complexity scaling and allowing cheap explicit computation of data reliability, i.e., the probability that the data comes from a definition domain of the learned manifold mapping. In comparison with the state-of-the-art linear and nonlinear methods on a set of noisy high-dimensional synthetic benchmarks, EOMC is demonstrated to provide a remarkably robust artefact-free learning and reconstruction of low-dimensional manifolds from noisy, nonlinear and nonstationary data. Application to the Lorenz-96 dynamical system in chaotic regime, as well as to a modified Hasegawa-Wakatani (mHW) model of drift-wave turbulence in the edge of a tokamak plasma reveals that for both of the models their essential dynamics is best described as a metastable regime-switching process, making infrequent transitions between the very persistent low-dimensional manifolds.  At the same time, the Markovian mean exit times and relaxation times (that bound the predictability horizons for the identified regime-switching process) appear to decrease only very slowly with the growing external forcing - indicating approximately two-fold longer prediction horizons than is currently anticipated based on analysis of positive Lyapunov exponents, even in very chaotic model regimes. It is also demonstrated that when applied for a lossy compression of the Lorenz-96 and mHW output data in various forcing regimes, EOMC achieves several orders of magnitude smaller compression loss - when compared to the common PCA-related linear compression approaches that build a backbone of the state-of-the-art lossy data compression tools (like JPEG, MP3, and others). These findings open new exciting opportunities for EOMC and transfer operator theory, by offering new possibilities to significantly improve predictive skills and performance of  data-driven tools in fluid  mechanics and geosciences applications.} 
\maketitle

\section{Introduction}
\label{sec:introduction}
Unsupervised dimensionality reduction and manifold learning are powerful techniques used to extract informative, low-dimensional representations from high-dimensional data, often for visualisation purposes, as a preprocessing step in machine learning pipelines, and in data compression algorithms \cite{Tenenbaum2000Global,Roweis2000Nonlinear,vanderMaaten2008Visualizing,Ma2011Manifold,Becht2018Evaluating,Sun2019Comparative}. Linear methods, such as Principal Component Analysis (PCA), assume that the data lies on or near a linear subspace, and analytically-exactly minimize the sum of squared Euclidean or kernelized  distances between the original full-dimensional data-points and their projections on a linear manifold \cite{Jolliffe2002PCA}. In contrast, nonlinear manifold learning algorithms like t-distributed Stochastic Neighbor Embedding (t-SNE), Uniform Manifold Approximation and Projection (UMAP), and methods based on diffusion and Laplacian eigenmaps ideas assume the data resides on a complex, curved manifold embedded within the high-dimensional space, and attempt to uncover this intricate local or global topological structure \cite{vanderMaaten2008Visualizing,majda12,McInnes2018,Healy2024UMAP}. While linear methods are often easier to interpret and apply, nonlinear methods are better at revealing complex patterns and clusters in data with nonlinear relationships, but can be more challenging to parameterize and interpret.

The primary distinction in computational scaling lies in the growth rate relative to the number of data points, \(T\), and the original dimensionality, \(D\). Principal Component Analysis (PCA) is a linear method that is highly efficient and typically scales as \(\mathcal{O}(TD^{2}+D^{3})\), or sometimes as \(\mathcal{O}(T^{2}D)\) depending on the implementation \cite{Jolliffe2002PCA}. This makes it very fast and suitable for large datasets. PCA intrinsically produces an explicit, linear mapping function (the principal components) which can be directly applied to project new, unseen data points into the low-dimensional space without re-training. Methods based on the eigenmap ideas and t-distributed Stochastic Neighbour Embedding (t-SNE) have a significantly higher base polynomial complexity due to the need to compute pairwise similarities in the high-dimensional space. The calculation of the initial distance matrix involves a complexity of \(\mathcal{O}(DT^{2})\)  \cite{majda12,vanderMaaten2008Visualizing}. The optimization step in the original algorithms scales around \(\mathcal{O}(T^{2})\) or even \(\mathcal{O}(T^{3})\). Optimized versions, like Barnes-Hut t-SNE, improve the optimization complexity to \(\mathcal{O}(DT\log(T))\) or \(\mathcal{O}(T\log(T))\) if the distance matrix is precomputed. Uniform Manifold Approximation and Projection (UMAP) was designed to be more computationally efficient than t-SNE and generally scales as \(\mathcal{O}(DT\log(T))\) or \(\mathcal{O}(T\log(T))\) in practice \cite{vanderMaaten2014Accelerating}. Regarding the explicit mapping function for unseen data points, eigenmap methods,  t-SNE and UMAP, in their standard, non-parametric forms, inherently do not yield a generalizable function. In contrast to the linear methods like PCA that provide explicit algebraic rules for manifold projection, nonlinear methods  typically only provide the specific embedding coordinates for the data points included in the training set. 

Researchers have developed various methods for out-of-sample (OOS) extension of nonlinear manifold learning methods, but these often introduce new computational deficits and complexities compared to PCA's simple linear projection:
\begin{itemize}
\item {\bf Interpolation Methods}: Projecting a single new point requires a search operation scaling as \(\mathcal{O}(DT)\) naively, or \(\mathcal{O}(D\log(T))\) using optimized search trees. The training is usually integrated into the original algorithm's runtime and does not add a separate substantial cost that in the best case remains  \(\mathcal{O}(DT\log(T))\) as in the baseline algorithm \cite{umap_interp}.
\item {\bf Kernel Methods}: These require an expensive training phase for Kernel PCA with a complexity of around \(\mathcal{O}(T^{3})\) for eigenvalue decomposition. Projecting a single new data point in inference scales as \(\mathcal{O}(DT)\) due to the need to compute kernel similarity against all training points  \cite{umap_kernel}.
\item {\bf Parametric Methods}: These involve training an auxiliary, explicit function (e.g., a neural network in Parametric UMAP) to map inputs to embeddings. The training complexity of such approaches is highly variable and depends on the network architecture, but it can be substantial. Once trained, however, the inference (projection) of a new point is very efficient, typically scaling linearly with dimensionality \(\mathcal{O}(D)\) for a simple forward pass. It is worth noting that more complex parametric architectures, such as the transformer models, intrinsically have a polynomial scaling for their core operations \cite{TabPFN25}. For instance, the self-attention mechanism in the original transformer model scales quadratically with the sequence length (which in a data context might relate to  \(T\) or a sequence of tokens representing a data point). This means that while they can potentially learn very complex nonlinear mappings, their training and application complexity remains polynomial, and often more computationally demanding than simpler neural network architectures used for Parametric UMAP, particularly as \(T\) or the input representation size grows  \cite{umap_param}.
\end{itemize}

Another class of nonlinear manifold learning methods is PCA-clustering, that exploits the idea of combining two linearly-scaling methods - clustering methods like K-means or Hidden Markov Models (HMMs) with PCA, exchanging the squared Euclidean distance in the K-means loss function with the squared Euclidean distance between the data points and their projections on cluster-specific linear manifolds \cite{horenko06,horenko08,metzner12}. As will be shown below on several examples, despite of its linear iteration complexity scaling in \(T\), PCA-clustering struggles to approximate nonlinear low-dimensional manifolds. 

In the following, we will start by briefly introducing the mathematical formulation of PCA-clustering (more details can be found in \cite{horenko06,horenko08,metzner12}), followed by demonstrating that the main bottleneck of PCA-clustering  is successfully mitigated without increasing the leading order of the cost scaling, by the two modifications of its mathematical formulation: (i) by upgrading the original manifold distance loss from PCA (being a semi-norm) to the weighted combination of manifold distance and squared Euclidean distances, hereby making the clustering loss function to a weighted norm (metrisation step); (ii) and by including the Shannon entropy regularization on the cluster affiliation probability measures, making the resulting clusterings entropy-optimal (entropy regularization step). Next, the resulting unsupervised Entropy Optimal Clustering algorithm (EOMC) will be investigated mathematically, and compared to the most popular linear and nonlinear manifold learning algorithms on synthetic benchmark examples, and on the outputs of the Lorenz-96 model of simplified atmospheric dynamics in chaotic and very chaotic regimes. Finally, the findings from Lorenz-96 model analysis will be validated applying EOMC to a much more advanced modified  Hasegawa-Wakatani (mHW) plasma turbulence model from Magnetohydrodynamics (MHD).     

\section{Methods}
\label{sec:methods}
\subsection{Mathematical formulation of the PCA-clustering}
\label{sec:m_formulation}
Let $X\in\mathbb{R}^{D,T}$ be a be a $(D\times T)$-dimensional real-valued data matrix, with every column  $X(:,t)$ representing a  $D$-dimensional vector of feature values for a data instance with an index $t$, where $t=1,\dots,T$ ($:$ denotes a column-extraction operation).   
 Let there be $K$ clusters, each of them is characterized by its centroid $\mu_k\in\mathbb{R}^{D}$  and the orthogonal $d$-dimensional linear manifold projector $\mathcal{T}_k\in\mathbb{R}^{D,d}$, $k=1,\dots,K$. In other words, $\mathcal{T}_k$ parameterizes the local linear approximation in a cluster $K$ and $\mathcal{T}_k$ is the associated linear and orthogonal projection operator. The $D$-dimensional reconstruction $X^{\textrm{rec},(k)}(:,t)$ of a given data-point  $X(:,t)$  after its projection on the linear manifold $k$ (defined by $\left\{\mu_k,\mathcal{T}_k\right\}$) is given as:  
\begin{eqnarray}
\label{eq:pca-clustering-reconst}
X^{\textrm{rec},(k)}(:,t)&=&\mu_k+\mathcal{T}_k\mathcal{T}_k^\dagger\left(X(:,t)-\mu_k\right),
\end{eqnarray}
where $\dagger$ denotes a matrix transposition operation. Then, the convex reconstruction $X^{\textrm{rec}}(:,t)$  of the original data point $X(:,t)$ from its projections on all of the cluster manifolds can be computed as a convex linear combination of individual reconstructions:
\begin{eqnarray}
\label{eq:pca-clustering-reconst_full}
X^{\textrm{rec}}(:,t)&=&\sum_{k=1}^K\gamma(k,t)X^{\textrm{rec},(k)}(:,t),
\end{eqnarray}
where
\begin{eqnarray}
\label{eq:pca-clustering-gamma}
\gamma(k,t)&\geq&0,\quad\textrm{and }\sum_{k=1}^K\gamma(k,t)=1,\quad \forall t,k.
\end{eqnarray}
Note that the $\gamma(k,t)$ may alternatively be interpreted as intrinsic/internal coordinates of
the manifold, or as probabilities of the data point $X(:,t)$ to be associated with the k-th cluster. For a fixed $K$  optimal values of $\gamma$ and  $\left\{\mu_k,\mathcal{T}_k\right\}$, $k=1,\dots,K$ can be found as a solution of the following minimization problem:
\begin{eqnarray}
\label{eq:pca-clustering_before_Jensen}
 &&\left\{\gamma^*,\mu^*_1,\mathcal{T}^*_1,\dots,\mu^*_K,\mathcal{T}^*_K\right\}=\arg\min\tilde{\mathcal{L}}\\
\label{eq:pca-clustering_before_Jensen2}
\tilde{\mathcal{L}}&=&\frac{1}{T}\sum_{t=1}^T\|X(:,t)-\sum_{k=1}^K\gamma(k,t)X^{\textrm{rec},(k)}(:,t)\|_2^2,\\
\label{eq:pca-clustering_before_Jensen3}
\textrm{s.t.}&&\mathcal{T}_k^\dagger\mathcal{T}_k=I_d, \\
\label{eq:pca-clustering_before_Jensen4}
&&\gamma(k,t)\geq0,\quad\textrm{and }\sum_{k=1}^K\gamma(k,t)=1,\quad \forall t,k.
\end{eqnarray}
Substituting (\ref{eq:pca-clustering-reconst}) in (\ref{eq:pca-clustering_before_Jensen2}) and applying the Jensen inequality, we obtain that the  values of $\gamma$ and  $\left\{\mu_k,\mathcal{T}_k\right\}$, $k=1,\dots,K$ can be approximated by minimizing the Jensen's upper bound $\mathcal{L}, \mathcal{L}\geq\tilde{\mathcal{L}}$ of the problem (\ref{eq:pca-clustering_before_Jensen}-\ref{eq:pca-clustering_before_Jensen4}):
\begin{eqnarray}
\label{eq:pca-clustering_after_Jensen}
 &&\left\{\gamma^*,\mu^*_1,\mathcal{T}^*_1,\dots,\mu^*_K,\mathcal{T}^*_K\right\}=\arg\min\mathcal{L}\\
\label{eq:pca-clustering_after_Jensen2}
\mathcal{L}&=&\frac{1}{T}\sum_{k=1}^K\sum_{t=1}^T\gamma(k,t)\|\left(X(:,t)-\mu_k\right) - \mathcal{T}_k\mathcal{T}_k^\dagger\left(X(:,t)-\mu_k\right)\|_2^2,\\
\label{eq:pca-clustering_after_Jensen3}
\textrm{s.t.}&&\mathcal{T}_k^\dagger\mathcal{T}_k=I_d, \\
\label{eq:pca-clustering_after_Jensen4}
&&\gamma(k,t)\geq0,\quad\textrm{and }\sum_{k=1}^K\gamma(k,t)=1,\quad \forall t,k.
\end{eqnarray}
PCA-clustering algorithm iteratively finds locally-optimal solutions of the problem (\ref{eq:pca-clustering_after_Jensen}-\ref{eq:pca-clustering_after_Jensen4}) using  two of its mathematical properties: (i) for a fixed $\gamma$, this problem has an analytic solution with respect to manifold parameters  $\left\{\mu_k,\mathcal{T}_k\right\}$, $k=1,\dots,K$, provided by the cluster-wise means and the dominant eigenvectors of the cluster-wise covariance matrices (i.e., means and covariances computed for each of the clusters separately); and (ii)  for the fixed manifold parameters  $\left\{\mu_k,\mathcal{T}_k\right\}$, $k=1,\dots,K$ and for each of the $t$, $\gamma(k^*,t)=1$ for $k^*=\arg\min_{k}\|\left(X(:,t)-\mu_k\right) - \mathcal{T}_k\mathcal{T}_k^\dagger\left(X(:,t)-\mu_k\right)\|_2^2$, and $\gamma(k,t)=0$ for all $k\neq k^*$ delivers the analytic $\gamma$-solutions of  (\ref{eq:pca-clustering_after_Jensen}-\ref{eq:pca-clustering_after_Jensen4}) (for details of the proof, see Step 3 of the Theorem proof in the supplement information of \cite{horenko_17a}) \footnote{Please note that since the optimal $\gamma$ in this solution takes only $0/1$-values, Jensen inequality becomes the Jensen equality and, hence, $\mathcal{L}=\tilde{\mathcal{L}}$}. PCA-clustering starts with a random intialization of $\gamma$ and iteratively repeats these analytic steps (i) and (ii), resulting in the monotonic convergence of $\mathcal{L}$, with overall iteration complexity scaling linearly in $T$  \cite{horenko06,horenko08,metzner12}. 

However, the Lemmas 1 and 2 below give rise to a significant mathematical difficulty, when applying PCA-clustering to approximation of low-dimensional ($D>d$) linear and nonlinear manifolds.

\begin{lemma}
Let $\mathcal{T}$ be a real-valued $D \times d$ matrix ($D > d$) with orthonormal columns (i.e., $\mathcal{T}^\dagger \mathcal{T} = I_d$, where $I_d$ be the $d \times d$ identity matrix). The kernel of the operator $B = I_D - \mathcal{T}\mathcal{T}^\dagger$ is non-empty; specifically, it contains non-zero vectors.
\end{lemma}

\begin{proof}
We analyze the operator $B = I_D - P$, where $P = \mathcal{T}\mathcal{T}^\dagger$ is the $D \times D$ orthogonal projection matrix onto the $d$-dimensional column space of $\mathcal{T}$, denoted $C(\mathcal{T})$.

To show that the kernel of $B$ is non-empty, we need to show that there exists at least one non-zero vector $v \in \mathbb{R}^D$ such that $Bv = 0$.
\begin{eqnarray}
Bv &=& (I_D - P)v = 0,\\
Iv - Pv &=& 0,\\
Pv &=& v.
\end{eqnarray}
Hence, any vector $v$ in the kernel of $B$ must be an eigenvector of $P$ corresponding to the eigenvalue $\lambda = 1$.

The set of all eigenvectors corresponding to $\lambda = 1$ forms the column space of $\mathcal{T}$, $C(\mathcal{T})$. Since $P$ is an orthogonal projection onto $C(\mathcal{T})$, any vector $v \in C(\mathcal{T})$ satisfies $Pv = v$. Because the matrix $\mathcal{T}$ has $d$ orthonormal columns, the dimension of $C(\mathcal{T})$ is exactly $d$.

The kernel of $B$ is the orthogonal complement of the column space, $C(\mathcal{T})^\perp$. The dimension of this kernel (called the nullity of $B$) is given by the rank-nullity theorem:
\begin{eqnarray}
\mathrm{dim}(\mathrm{Ker}(B)) = D - \mathrm{rank}(P) = D - d.
\end{eqnarray}
Since the problem statement assumes $D > d$, the dimension of the kernel $D-d$ is a positive integer (at least 1).

A subspace with a positive dimension contains non-zero vectors. For instance, any non-zero vector that is orthogonal to every column of $\mathcal{T}$ will be in the kernel of $B$.

Therefore, the kernel of $B = I_D - \mathcal{T}\mathcal{T}^\dagger$ is non-empty and contains non-zero vectors.
\end{proof}

\begin{lemma}
\label{lemma:one}
Let $x\in\mathbf{R}^D$. Then, for any $\xi\in\textrm{Ker}\left(I_D-\mathcal{T}_k\mathcal{T}_k^\dagger\right)$  and $x^\xi=x+\xi$ it holds that $\|\left(x-\mu_k\right) - \mathcal{T}_k\mathcal{T}_k^\dagger\left(x-\mu_k\right)\|_2^2=\|\left(x^\xi-\mu_k\right) - \mathcal{T}_k\mathcal{T}_k^\dagger\left(x^\xi-\mu_k\right)\|_2^2$.
\end{lemma}
\begin{proof} 
\begin{eqnarray}
\label{eq:lemma-proof}
 &&\|\left(x^\xi-\mu_k\right) - \mathcal{T}_k\mathcal{T}_k^\dagger\left(x^\xi-\mu_k\right)\|_2^2=\|\left(I_D-\mathcal{T}_k\mathcal{T}_k^\dagger\right)\left(x^\xi-\mu_k\right)\|_2^2=\nonumber\\
 &&=\|\left(I_D-\mathcal{T}_k\mathcal{T}_k^\dagger\right)\left(x-\mu_k\right)+\underbrace{\left(I_D-\mathcal{T}_k\mathcal{T}_k^\dagger\right)\xi}_{\xi\in\textrm{Ker}\left(I_D-\mathcal{T}_k\mathcal{T}_k^\dagger\right)}\|_2^2=\|\left(I_D-\mathcal{T}_k\mathcal{T}_k^\dagger\right)\left(x-\mu_k\right)+0\|_2^2=\nonumber\\
&&=\|\left(I_D-\mathcal{T}_k\mathcal{T}_k^\dagger\right)\left(x-\mu_k\right)\|_2^2,
\end{eqnarray}
where $I_D$ be the $D \times D$ identity matrix.
\end{proof} 

From Lemmas 1 and 2 follows the first major problem of PCA-clustering: according to the Lemmas, the kernel of the projection operator is non-empty when $d<D$ (Lemma 1), and any perturbations inside of the kernel  do not have any effect on the loss (Lemma 2).  Hence, loss function in (\ref{eq:pca-clustering_after_Jensen}-\ref{eq:pca-clustering_after_Jensen4}) is a semi-norm, not penalizing errors in the directions that are orthogonal, or close to orthogonal with respect to the manifolds $\mathcal{T}_k$.  As will be demonstrated on examples below (see, e.g., Fig.~2C and the left panel of Fig.~3C), this can lead to the PCA-clustering minimizers $\mathcal{T}_k$ that become orthogonal, or close to orthogonal, with respect to the actual nonlinear low-dimensional manifold - hereby ignoring the actual manifold data, and achieving a small value of $\mathcal{L}$ at the same time. The second problem of PCA-clustering is induced by the exact $0/1$ minimizer in the step (ii) described above: it confines the resulting manifold reconstruction $X^{\textrm{rec}}(:,t)$ in (\ref{eq:pca-clustering-reconst_full}) to piece-wise linear functions only, not allowing to achieve smooth convex interpolations of nonlinear manifolds that should be possible with the original optimization problem formulation (\ref{eq:pca-clustering_before_Jensen}-\ref{eq:pca-clustering_before_Jensen4}). But, this original formulation  (\ref{eq:pca-clustering_before_Jensen}-\ref{eq:pca-clustering_before_Jensen4}) does not allow obtaining the cheaply-computable analytic linearly-scalable solutions (i)-(ii) that are used in the PCA-clustering algorithm. And, solving the original problem (\ref{eq:pca-clustering_before_Jensen}-\ref{eq:pca-clustering_before_Jensen4}) - being highly-nonlinear, subject to multivariate and polynomial constraints - with standard numerical tools from optimization theory would require very expensive numerical algorithms that  have the exponential worst-case scaling \cite{qp_np}.     

\subsection{Entropy-Optimal Manifold Clustering (EOMC)}
\label{sec:eomc}
To mitigate these two central problems of PCA-clustering, the two following modifications of the loss function $\mathcal{L}$ in (\ref{eq:pca-clustering_after_Jensen}-\ref{eq:pca-clustering_after_Jensen4}) will be adopted:
\begin{itemize}
\item {\bf Loss metrisation}: to fix the problem identified by the Lemmas 1 and 2, we will introduce an additional scalar hyper-parameter $\alpha\geq 0$ and add an addtional term $\alpha\|X(:,t)-\mu_k\|_2^2$ to the loss of each of the $t$ instances for all $k=1,\dots,K$. As will be proven below, this modification upgrades the semi-norm loss of (\ref{eq:pca-clustering_after_Jensen}-\ref{eq:pca-clustering_after_Jensen4}) to the $\alpha$-weighted norm, fixing the problem with the data points  in the directions that are close to orthogonal to $\mathcal{T}_k$. 
\item {\bf Entropy regularization}: to avoid the problem with a piece-wise linearity restriction - and  to allow for the $\gamma$-solutions to be not only $0/1$  (and, to achieve this without returning back to the original formulation (\ref{eq:pca-clustering_before_Jensen}-\ref{eq:pca-clustering_before_Jensen4})  that would result in exponentialy-scaling worst case numerics) - we can use the fact that the columns of  $\gamma$ can be interpreted as $t$-dependent $K$-dimensional probability measures.  We can  add the normalized negative Shannon entropy regularization terms  $\beta\gamma(k,t)\textrm{log}_K\left(\gamma(k,t)\right)$ (with a  scalar hyper-parameter $\beta\geq 0$) to the loss, to optimally-adjust the entropies of these probability measures during learning\footnote{Base $K$ of the logarithm is used to normilize the entropy values to the interval $\left[0,1\right]$}. This modification is motivated by the entropic learning methods that allow training the \emph{entropy-optimal} probability measures by tuning this hyper-parameter $\beta$ in the loss. The $0/1$ $\gamma$-solution of the PCA-clustering is a minimum entropy measure, and is attained when setting $\beta=0$ \cite{Horenko_2020,horenko_pnas_22,espa_22,horenko_pnas_23,bassetti25}.
\end{itemize}
Making these two modifications of PCA-clustering loss and re-arranging the terms we obtain:
\begin{eqnarray}
\label{eq:eomc}
 &&\left\{\gamma^*,\mu^*_1,\mathcal{T}^*_1,\dots,\mu^*_K,\mathcal{T}^*_K\right\}=\arg\min\mathcal{L}^{\textrm{EOMC}},\\
\label{eq:eomc2}
\mathcal{L}^{\textrm{EOMC}}&=&\frac{1}{T}\sum_{k,t=1}^{K,T}\gamma(k,t)\left[\left(X(:,t)-\mu_k\right)^\dagger\left(\left(I+\alpha\right) - \mathcal{T}_k\mathcal{T}_k^\dagger\right)\left(X(:,t)-\mu_k\right)+\beta\textrm{log}_K\left(\gamma(k,t)\right)\right],\\
\label{eq:eomc3}
\textrm{s.t.}&&\mathcal{T}_k^\dagger\mathcal{T}_k=I_d, \\
\label{eq:eomc4}
&&\gamma(k,t)\geq0,\quad\textrm{and }\sum_{k=1}^K\gamma(k,t)=1,\quad \forall t,k.
\end{eqnarray}

Please note that the PCA-clustering \cite{horenko06,horenko08,metzner12} is  a special case of  (\ref{eq:eomc}-\ref{eq:eomc4}), attained when setting $\alpha=0$ and $\beta=0$. Standard linear PCA is also a special case of (\ref{eq:eomc}-\ref{eq:eomc4}), when selecting $K=1, \alpha=0, \beta=0$.

Before we continue, we will need to formulate and to prove two auxiliary Lemmas 3 and 4, that would in the following help us to mitigate the problem induced by the Lemmas 1 and 2 above.
\begin{lemma}
Let $\mathcal{T}$ be a real-valued $D \times d$ matrix  with orthonormal columns (i.e., $\mathcal{T}^\dagger \mathcal{T} = I_d$, where $I_d$ is the $d \times d$ identity matrix). Let $I_D$ be the $D \times D$ identity matrix, and let $\alpha > 0$ be a positive scalar. Then, the kernel of the operator $A^\alpha = (1+\alpha)I_D - \mathcal{T}\mathcal{T}^\dagger$ contains only the zero vector, and operator $A^\alpha$ is invertible.
\end{lemma}

\begin{proof}
We analyze the properties of the operator $A^\alpha = (1+\alpha)I_D - P$, where $P = \mathcal{T}\mathcal{T}^\dagger$.

The matrix $P$ is a $D \times D$ orthogonal projection matrix onto the $d$-dimensional column space of $\mathcal{T}$. A key property of any orthogonal projection matrix is that its only possible eigenvalues are $0$ and $1$.

To find the kernel of $A^\alpha$, we seek all vectors $v \in \mathbb{R}^D$ such that $A^\alpha v = 0$:
\begin{eqnarray}
A^\alpha v &=& ((1+\alpha)I_D - P)v = 0.
\end{eqnarray}
Expanding the equation gives:
\begin{eqnarray}
(1+\alpha)v - Pv = 0,\\
Pv = (1+\alpha)v.
\end{eqnarray}

This equation is an eigenvalue problem for the matrix $P$. For a non-zero vector $v$ to exist as a solution, the scalar $(1+\alpha)$ must be an eigenvalue of $P$.

However, we are given that $\alpha > 0$. Therefore, the proposed eigenvalue $(1+\alpha)$ satisfies:
\begin{eqnarray}
1+\alpha &>&1.
\end{eqnarray}

Since the set of all possible eigenvalues for $P$ is restricted to $\{0, 1\}$, and $1+\alpha$ is strictly greater than 1, $(1+\alpha)$ cannot be an eigenvalue of $P$.
Consequently, the equation $Pv = (1+\alpha)v$ has only one solution: the zero vector, $v = 0$.

Thus, the kernel of $A^\alpha$ contains only the zero vector, which implies that the operator $A^\alpha$ is invertible.
\end{proof}
\begin{lemma}
Let $x$ be a column vector in $\mathbb{R}^D$, and let $\mathcal{T}$ be a $D \times d$ orthogonal real valued matrix, with $d$ columns forming an orthonormal basis, such that $\mathcal{T}^\dagger \mathcal{T} = I_d$. Let $I_D$ be the $D \times D$ identity matrix, and $\alpha > 0$ be a positive scalar.

Then, the quadratic form defined by the function $f(x) = x^\dagger A^\alpha x$, where $A^\alpha = (1+\alpha)I_D - \mathcal{T}\mathcal{T}^\dagger$, is a strictly-convex function with a unique minimum achieved at $x = 0$.
\end{lemma}
\begin{proof}
A function $f(x) = x^\dagger A^\alpha x$ is strictly convex if and only if the matrix $A^\alpha$ is positive definite ($A^\alpha \succ 0$). A unique minimum exists if the function is strictly convex and the domain is $\mathbb{R}^D$, with the minimum occurring where the gradient is zero ($\nabla f(x) = 2A^\alpha x = 0$).

We analyze the matrix $A = (1+\alpha)I_D - \mathcal{T}\mathcal{T}^\dagger$. Let $P = \mathcal{T}\mathcal{T}^\dagger$ be the $D \times D$ orthogonal projection matrix.

To show that $A^\alpha$ is positive definite, we must show that $x^\dagger A^\alpha x > 0$ for all non-zero vectors $x \in \mathbb{R}^D$:
\begin{eqnarray}
x^\dagger A^\alpha x &=& x^\dagger \left( (1+\alpha)I_D - P \right) x = (1+\alpha)x^\dagger I_D x - x^\dagger P x = (1+\alpha)\|x\|^2 - \|Px\|^2.
\end{eqnarray}

Here, $\|x\|^2$ is the squared Euclidean norm of $x$. The term $P x$ represents the orthogonal projection of $x$ onto the $d$-dimensional column space of $\mathcal{T}$, $C(\mathcal{T})$. By the properties of orthogonal projections, the norm of the projected vector is always less than or equal to the norm of the original vector: $\|Px\|^2 \leq \|x\|^2$.

We can rewrite the expression as:
\begin{eqnarray}
x^\dagger A^\alpha x &=& \|x\|^2 + \alpha\|x\|^2 - \|Px\|^2 = (\|x\|^2 - \|Px\|^2) + \alpha\|x\|^2.
\end{eqnarray}

Since $\|Px\|^2 \leq \|x\|^2$, the first term $(\|x\|^2 - \|Px\|^2)$ is non-negative. Since $\alpha > 0$, the second term $\alpha\|x\|^2$ is strictly positive for any $x \neq 0$.

Therefore, $x^\dagger A^\alpha x > 0$ for all $x \neq 0$, which proves that the matrix $A^\alpha$ is positive definite.

Because $A^\alpha$ is positive definite, the function $f(x) = x^\dagger A^\alpha x$ is strictly convex. A strictly convex function defined on $\mathbb{R}^D$ has a unique global minimum. The gradient of $f(x)$ is $\nabla f(x) = 2A^\alpha x$. Setting the gradient to zero gives $2A^\alpha x = 0$. Since $A^\alpha$ is invertible (as proven in the Lemma 2), the only solution is $x=0$.

Thus, the strictly-convex function $f(x)$ has a unique minimum at $x=0$.
\end{proof}

Finally, as proven in the Theorem 1 below, optimization problem (\ref{eq:eomc}-\ref{eq:eomc4}) can be solved through a sequence of analytically-solvable steps - and without exceeding the computational cost of the original PCA-clustering algorithm for (\ref{eq:pca-clustering_after_Jensen}-\ref{eq:pca-clustering_after_Jensen4}), i.e.,  with the overall leading iteration cost scaling of $\mathcal{O}(T)$. 

\begin{theorem}
\label{theorem:theone}
Let $X\in\mathbb{R}^{D,T}$, and the hyper-parameters $K\geq1$, $d\leq K$, $\alpha> 0$  and $\beta\geq 0$ are fixed. Then, the EOMC problem (\ref{eq:eomc}-\ref{eq:eomc4}) has the following properties: 
\begin{enumerate}[(1.)]
\item \underline{Step 1 of EOMC-algorithm, analytic solutions for $\left\{\mu_k,\mathcal{T}_k\right\}$, $k=1,\dots,K$}:  if $\sum_{t=1}^T\gamma(k,t)>0$,  than  for a fixed $\gamma$  that satisfies the constraints  (\ref{eq:eomc4}), the solution of (\ref{eq:eomc}-\ref{eq:eomc4})  is provided by: 
\begin{eqnarray}
\label{eq:mu}
\mu_k^* &=&\frac{\sum_{t=1}^T\gamma(k,t)X(k,t)}{\sum_{t=1}^T\gamma(k,t)},\\
\label{eq:T}
\mathcal{T}_k^* &=&\underset{d}{\overline{\mathrm{eigvec}}}\left(\mathrm{Cov}_k\left(X,\gamma,\mu_k^*\right)\right),
\end{eqnarray}
where $\underset{d}{\overline{\mathrm{eigvec}}}\left(A\right)$ denotes an operation of computing $d$ dominant eigenvectors of A (i.e., the eigenvectors that correspond to the $d$ largest eigenvalues of the symmetric positive-semidefinite operator $A$) and whose $d$ columns consist of  $D$-dimensional mutually-orthogonal vectors\footnote{There is one issue here: suppose eigenvalues $\lambda_1\geq\dots\geq\lambda_{d-1}=\lambda_d=\lambda_{d+1}>\lambda_{d+2\geq\dots}$. In that case, not all of the eigenvectors associated with $\lambda_j$ for $j\in\{d-1,d,d+1\}$  are included in the result of $\underset{d}{\overline{\mathrm{eigvec}}}\left(A\right)$  operator. In such practical situations – that  are easiliy-detectable in real applications  (one just needs to inspect the spectrum) – one should increase the user-defined reduced dimension $d$ accordingly, such that all of the relevant dimensions corresponding to this multiple eigenvalue are included. } If  there exists  index $k$ such that $\sum_{t=1}^T\gamma(k,t)=0$, than any values of $\mu_k^*$ and any orthonormal $\mathcal{T}_k^*$ are a solution, the cluster is empty,  can be discarded and the hyperparameter $K$ can be reduced by one.

Operator $\mathrm{Cov}_k\left(X,\gamma,\mu_k^*\right)$ is defined as  $\mathrm{Cov}_k\left(X,\gamma,\mu_k^*\right)=\frac{\sum_{t=1}^T\gamma(k,t)\left(X(:,t)-\mu_k^*\right)\left(X(:,t)-\mu_k^*\right)^\dagger}{\sum_{t=1}^T\gamma(k,t)}$ (where $k$ is such that $\sum_{t=1}^T\gamma(k,t)>0$). Cost of (\ref{eq:mu}-\ref{eq:T}) computation scales as $\mathcal{O}\left(TKD\left(D+1\right)+KdD^2\right)$.
\item \underline{Step 2 of EOMC-algorithm, analytic solutions for $\gamma(k,t)$}:  let $g(t)=\left(g_1(t),\dots,g_K(t)\right)$, where $g_k(t)=\left(X(:,t)-\mu_k\right)^\dagger\left(\left(I+\alpha\right) - \mathcal{T}_k\mathcal{T}_k^\dagger\right)\left(X(:,t)-\mu_k\right)$, and let $j^*=\text{argmin}_{k}g_k(t)$ be the argument of the infimum for  a vector $g(t)$ with a fixed $t$. Then, for fixed $\left\{\mu_k,\mathcal{T}_k\right\}$, $k=1,\dots,K$ and when $\beta>0$, the solution of (\ref{eq:eomc}-\ref{eq:eomc4})  is given by: 
\begin{eqnarray}
\label{eq:gamma1}
\gamma^*(k,t) &=&\frac{\exp\left(-\beta^{-1}\left(g_k(t)-g_{j^*}(t)\right)\right)}{\sum_{k=1}^K\exp\left(-\beta^{-1}\left(g_k(t)-g_{j^*}(t)\right)\right)}.
\end{eqnarray}
If $\beta=0$, than the  solution of (\ref{eq:eomc}-\ref{eq:eomc4})  is given by $\gamma^*(k^*,t)=1$ for $k^*=\arg\min_{k}\left(g_k(t)-g_{j^*}(t)\right)$, and $\gamma^*(k,t)=0$ for all $k\neq k^*$. Cost of computing $\gamma^*$ with (\ref{eq:mu}-\ref{eq:T}) scales as $\mathcal{O}\left(TK\left(D+1\right)\right)$.
\item \underline{Monotonicity and convergence of EOMC-algorithm}: starting with a randomly-generated $\gamma$ that satisfies constraints in  (\ref{eq:eomc3}-\ref{eq:eomc4}), and iteratively repeating the above \underline{Step 1} and \underline{Step 2}, results in monotonic decrease of the function $\mathcal{L}^{\textrm{EOMC}}$, converging to a local optimum of  (\ref{eq:eomc}-\ref{eq:eomc4}). The overall computational iteration cost of EOMC-algorithm scales as $\mathcal{O}\left(TK\left(D+1\right)^2+KdD^2\right)$   
\item \underline{Projecting a new data point on the EOMC-manifold}: for $\beta>0$, nonlinear projection $Y^{\textrm{proj}}$ of any new $Y\in\mathcal{R}^D$ on the $d$-dimensional EOMC manifold defined by  $\left\{\mu_k,\mathcal{T}_k\right\}$, $k=1,\dots,K$, can be computed explicitly as: 
\begin{eqnarray}
\label{eq:projectionl}
Y^{\textrm{proj}}&=&\frac{\sum_{k=1}^K\exp\left(-\beta^{-1}\left(g_k-g_{j^*}\right)\right)\left(\mu_k+\mathcal{T}_k\mathcal{T}_k^\dagger\left(Y-\mu_k\right)\right),}{\sum_{k=1}^K\exp\left(-\beta^{-1}\left(g_k-g_{j^*}\right)\right)}
\end{eqnarray}
where  $g_k=\left(Y-\mu_k\right)^\dagger\left(\left(I+\alpha\right) - \mathcal{T}_k\mathcal{T}_k^\dagger\right)\left(Y-\mu_k\right)$, and $g_{j^*}$ is the minimal value attained by the $K$ elements of vector $g$, and $k=1,\dots,K$. If $\beta=0$, projection is computed as $Y^{\textrm{proj}}=\mu_{k^*}+\mathcal{T}_{k^*}\mathcal{T}_{k^*}^\dagger\left(Y-\mu_{k^*}\right)$, where $k^*=\arg\min_{k}\left(g_k(t)-g_{j*}(t)\right)$. This computation in both situations requires at most $\mathcal{O}\left(K\left(D+1\right)\right)$ operations. 
\end{enumerate}
\end{theorem}

\begin{proof}
\begin{enumerate}[(1.)]
\item Let $X \in \mathbb{R}^{D,T}, \gamma, K \geq 1, d, \alpha > 0$, and $\beta \geq 0$ be fixed. Then, according to Lemma 4, and for $k = 1, \dots, K$, the problem(\ref{eq:eomc}-\ref{eq:eomc4}) is a strictly convex optimization problem for finding $\mu_k^*$, such that
\begin{eqnarray}
&&\sum_{t=1}^T\gamma(t,k)\left(X(:,t)-\mu_k^*\right)=0,
\end{eqnarray}
that after re-arranging of terms provides (\ref{eq:mu})  if $\sum_{t=1}^T\gamma(k,t)>0$. If  $\sum_{t=1}^T\gamma(k,t)=0$, than any value of $\mu_k^*$ is a solution.
Next, we fix these $\mu_k^*$, for all $k=1,\dots,K$, and consider the solution of   (\ref{eq:eomc}-\ref{eq:eomc4}) with respect to $\mathcal{T}_k$, for fixed  $X\in\mathbb{R}^{D,T}$, $\gamma$ and the hyper-parameters $K\geq1$, $d\leq K$, $\alpha> 0$  and $\beta\geq 0$. Defining the $d\times d$ matrices of Lagrange multipliers $\Lambda^(k)$, we can rewrite (\ref{eq:eomc}-\ref{eq:eomc4}) in the Euler-Lagrange form
 \begin{eqnarray}
\label{eq:euler}
\tilde{\mathcal{L}}_\Lambda^{\textrm{EOMC}}&=&\frac{1}{T}\sum_{k=1}^K\sum_{t=1}^T\gamma(k,t)\left[\left(X(:,t)-\mu_k^*\right)^\dagger\left(\left(I+\alpha\right) - \mathcal{T}_k\mathcal{T}_k^\dagger\right)\left(X(:,t)-\mu_k^*\right)\right]+\nonumber\\
&&+\sum_{k=1}^K\sum_{i,j=1}^d\Lambda_{i,j}^{(k)}\left(\left\{I_d\right\}_{i,j}-\mathcal{T}_k^\dagger(:,i)\mathcal{T}_k(:,j)\right).
 \end{eqnarray}
Taking the gradients of $\tilde{\mathcal{L}}_\Lambda^{\textrm{EOMC}}$ with respect to $ \mathcal{T}_k$ and $\Lambda^{(k)}$, setting them to zero and re-arranging terms results in the following system of equations   
 \begin{eqnarray}
\label{eq:cov}
\mathrm{Cov}_k\left(X,\gamma,\mu_k^*\right)\mathcal{T}_k&=&\Lambda^{(k)}\mathcal{T}_k,\\
\mathcal{T}_k^\dagger\mathcal{T}_k&=&I_d, \nonumber
 \end{eqnarray}
 for all $k=1,\dots,K$. Substituting (\ref{eq:cov}) into (\ref{eq:euler}), taking trace and deploying algebraic transformations that make use of the trace operation properties, we obtain that $\Lambda^{(k)}$ in (\ref{eq:cov}) is a diagonal matrix of dominant eigenvalues of $\mathrm{Cov}_k\left(X,\gamma,\mu_k^*\right)$ (trace of $\Lambda^{(k)}$ should  maximize the trace of  $\mathrm{Cov}_k\left(X,\gamma,\mu_k^*\right)$ after projection on the orthogonal subspace $ \mathcal{T}_k$). Hence, $ \mathcal{T}_k$ corresponds column-wise to the $d$ dominant eigenvectors of   $\mathrm{Cov}_k\left(X,\gamma,\mu_k^*\right)$  and $\mathcal{T}_k^* =\underset{d}{\overline{\mathrm{eigvec}}}\left(\mathrm{Cov}_k\left(X,\gamma,\mu_k^*\right)\right)$. Summing up the cost of computing the weighted means and weighted covariances for each cluster with the cost of computing the $d$-dominant eigenvectors of symmetric positive-semidefinite matrix  $\mathrm{Cov}_k\left(X,\gamma,\mu_k^*\right)$ (that is less than a cost of full diagonalization, and allows applying iterative Krylov methods with cost scaling of $\mathcal{O}\left(KdD^2\right)$), we obtain the cost scaling of  $\mathcal{O}\left(TKD\left(D+1\right)+KdD^2\right)$ for \underline{Step 1} .

\item These properties follow from applying  Lemma 2.1 and Lemma 2.6 in \cite{bassetti25} to  (\ref{eq:eomc}-\ref{eq:eomc4}). The cost of computing \underline{Step 2}  consists of $\mathcal{O}\left(TKD\right)$ (computing $K$ elements of vector $g(t)$ for all $t$), and adding the costs of $\mathcal{O}\left(TK\right)$ computations of (\ref{eq:gamma1}) for every $t$. 
\item As follows from the above proofs for items (1.) and (2.), each of the \underline{Step 1} and \underline{Step 2} leads to a monotonic decrease of the function value $\mathcal{L}^{\textrm{EOMC}}$ in (\ref{eq:eomc}-\ref{eq:eomc4}). It is straightforward to validate that this sequence is bounded from below with $\mathcal{L}^{\textrm{EOMC}}\geq -\beta$.  Hence, since iterative repetition of these steps also generates a monotonically-decreasing sequence - and it is bounded from below with $-\beta$ in the space of real numbers $\mathcal{R}^1$ - this monotonically-decreasing sequence is converging to some finite $\mathcal{L}^{\textrm{EOMC}}$. 
\item These properties and cost scaling follow directly from applying property (2.) of the Theorem to a new data point $Y\in\mathcal{R}^D$. 
\end{enumerate}
\end{proof}

\subsection{Incorporating simultaneous learning of data reliability in EOMC}
\label{sec:gamma_0}
Learning piece-wise low-dimensional linear manifold approximations with EOMC implicitly deploys an assumption that all of the data instances have the same uniform probability $(1/T)$ to contribute to the loss function $\mathcal{L}^{\textrm{EOMC}}$ in (\ref{eq:eomc}-\ref{eq:eomc4}). This is not a problem when this assumption is justified, and when the training data is spread more-or-less uniformly and close to the manifold. But, this becomes a problem when the new data instances $X$ (f.e., the ones that were not used in the training), come from somewhere arbitrarily in the feature space - and further away from the manifold. In the Sec.\ref{sec:synt_ex3} below, we provide an example of such a situation - resulting in projection artefacts, when training the manifold with the data close to the manifold (red crosses in Fig.~4), and then using the trained EOMC manifold projectors (\ref{eq:projectionl}) from Theorem 1 for the data points that are far away from the manifold and from the training data (blue dots in Fig.~4B). This \emph{input data reliability issue} is a very general and severe problem in contemporary AI, resulting in low robustness, "over-confidence" in provided answers, and in AI vulnerability  to the so-called adversarial attacks, when subject to new input data that is sufficiently-different from the domain $\Omega$ of "admissible" data implicitly defined by the training data.

To address this issue - and following the ideas originally introduced in the Entropic Outlier Sparsification (EOS) and Entropy-Optimal Networks (EON) methods \cite{horenko_pnas_22,bassetti25}  - we propose to incorporate in the EOMC problem formulation (\ref{eq:eomc}-\ref{eq:eomc4}) an additional \emph{learnable} variable $\gamma_0(t)=\bP\left[X(:,t)\in\Omega\right]$,  measuring a probability that a new data instance $X(:,t)$  belongs to the "admissible" data domain $\Omega$. Denoting $l(t)=-\sum_{k=1}^K\gamma(k,t)\left[\left(X(:,t)-\mu_k\right)^\dagger\left(\left(I+\alpha\right) - \mathcal{T}_k\mathcal{T}_k^\dagger\right)\left(X(:,t)-\mu_k\right)+\beta\textrm{log}_K\left(\gamma(k,t)\right)\right]$, we can now rewrite (\ref{eq:eomc}-\ref{eq:eomc4}) as a maximisation of the expectation $\bE_{\gamma_0}\left[l\right]$ of the loss $l$ wrt. this probability measure $\gamma_0$, subject to the Shannon entropy regularization terms  $\beta_0\gamma_0\left(t\right)\textrm{log}_T\gamma_0\left(t\right)$, where $\beta_0\geq0$. Re-arranging terms we obtain:       
\begin{eqnarray}
\label{eq:eomc_g0}
 &&\left\{\gamma_0^*,\gamma^*,\mu^*_1,\mathcal{T}^*_1,\dots,\mu^*_K,\mathcal{T}^*_K\right\}=\arg\min\mathcal{L}_{\gamma_0}^{\textrm{EOMC}}\\
\label{eq:eomc_g02}
\mathcal{L}_{\gamma_0}^{\textrm{EOMC}}&=&\sum_{t=1}^T\gamma_0(t)\left[\beta_0\textrm{log}_T\gamma_0\left(t\right)-l(t)\right],\\
\textrm{s.t.}&&l(t)=-\sum_{k=1}^K\gamma(k,t)\left[\left(X(:,t)-\mu_k\right)^\dagger\left(\left(I+\alpha\right) - \mathcal{T}_k\mathcal{T}_k^\dagger\right)\left(X(:,t)-\mu_k\right)+\beta\textrm{log}_K\left(\gamma(k,t)\right)\right], \nonumber\\
\label{eq:eomc_g03}
&&\\
\label{eq:eomc_g04}
&& \mathcal{T}_k^\dagger\mathcal{T}_k=I_d, \\
\label{eq:eomc_g05}
&&\gamma(k,t)\geq0,\quad\textrm{and }\sum_{k=1}^K\gamma(k,t)=1,\quad \forall t,k,\\
\label{eq:eomc_g06}
&&\gamma_0(t)\geq0,\quad\textrm{and }\sum_{t=1}^T\gamma_0(t)=1,\quad \forall t.
\end{eqnarray}
Shannon entropy regularization terms  $\beta_0\gamma_0\left(t\right)\textrm{log}_T\gamma_0\left(t\right)$ are incorporated here with a positive sign, since for $\beta_0\equiv0$, $\gamma_0$ solution of (\ref{eq:eomc_g0}-\ref{eq:eomc_g06}) is \emph{entropy-minimal}, attaining only $1/0$ values (see, e.g.,  Step 3 of the Theorem proof in the supplement information of \cite{horenko_17a} or Lemma 2.1 from \cite{bassetti25} for a formal proof) - and the \emph{entropy-optimal} $\gamma_0$-solutions can only be found when \emph{maximizing} the Shannon entropy. It is also straightforward to validate that the previous formulation (\ref{eq:eomc}-\ref{eq:eomc4}) becomes a special case of  (\ref{eq:eomc_g0}-\ref{eq:eomc_g06}) when $\beta_0\rightarrow+\infty$ (since then $\gamma_0(t)\rightarrow (1/T)$, for all $t$).

Then, following the same line of argumentation as in the Step 2 proof of Theorem 1 above (applying Lemma 2.6 \cite{bassetti25}), it is easy to show that for the fixed values of the loss $l$ (hence, when all other learning variables except of $\gamma_0$ are fixed), and for any $\beta_0>0$, there exists a unique solution $\gamma^*_0$ of the problem (\ref{eq:eomc_g0}-\ref{eq:eomc_g06}):
\begin{eqnarray}
\label{eq:gamma0}
\gamma_0^*(t) &=&\frac{\exp\left(\beta_0^{-1}\left(l_{\max}+l(t)\right)\right)}{\sum_{t=1}^T\exp\left(\beta_0^{-1}\left(l_{\max}+l(t)\right)\right)},
\end{eqnarray}
 where  $l_{\max}$ is the maximum of $l$ over all $t$. This solution can be computed with an additional cost of $\mathcal{O}(T)$ - independent of $K$ and $D$ (since the values of loss $l$ are pre-computed in the original EOMC algorithm and can be recycled). Hence, adding \underline{Step 3} (\ref{eq:gamma0}) to the original algorithm described in the Theorem 1 above can be done essentially for free, not increasing the leading order of the EOMC algorithm complexity, and not violating the monotonicity of EOMC convergence. 

\subsection{Selection of EOMC hyper-parameters $d, K, \alpha, \beta, \beta_0$}
\label{sec:hyper}
In comparison with the state-of-the-art nonlinear manifold learning methods like t-SNE and UMAP, EOMC relies on a  smaller set of hyper-parameters. It requires to tune the reduced manifold dimensionality $d$, the number $K$ of locally-linear manifolds for nonlinear approximation, as well as two or three non-negative scalar regularization parameters $\alpha$, $\beta$ (and $\beta_0$, if simultaneous learning of data reliability is incorporated in EOMC, as described in Sec.~\ref{sec:gamma_0}). 

Methods like t-SNE require tuning a much larger set of hyper-parameters, that besides of the reduced manifold dimensionality parameter $d$ and number of PCA dimensions $K$, include such model-specific adjustable parameters like perplexity, exaggeration, and Barnes-Hut tradeoff parameter. In addition, UMAP model-specific hyper-parameters include  the number of neighbours, metric and its weight, and minimal distance between embedded points. Moreover, numerics of these nonlinear manifold learning methods relies on (stochastic) gradient descent algorithm - that requires tuning of multiple additional hyper-parameters that can have a very strong effect 
 on convergence and cost of the learning phase (like learning rate schedule, batch size, cache size, etc.).  
 
 As shown in the Theorem 1 above, EOMC is performed without deploying the (stochastic) gradient numerics. Instead, EOMC performs an iterative repetition of analytic solutions not requiring additional numerics-specific hyper-parameter tuning in  \underline{Step 1},  \underline{Step 2} (and in additional \underline{Step 3} defined by (\ref{eq:gamma0}), if data reliability is also learned).   

As will be demonstrated below on practical examples (for example, see Fig.~1), tuning the  EOMC hyper-parameters $d$ and $K$ can be done iteratively. To select optimal values for  $d$ and $K$, for each of the $k=1,\dots,K$ manifolds  one inspects the decay of eigenvalues $\Lambda^{(k)}_{1,1},\Lambda^{(k)}_{2,2},\dots,\Lambda^{(k)}_{D,D}$ for the local weighted covariances $\mathrm{Cov}_k\left(X,\gamma,\mu_k^*\right)$ in the local manifolds: as follows from the Theorem 1 above, $\sum_{i=d+1}^D\Lambda^{(k)}_{i,i}$ quantifies the amount of squared 2-norm loss from projecting the data, that is "assigned" to this manifold $k$, through the EOMC internal coordinates $\gamma(k,:)$. Hence, it quantifies the local quality of lossy compression, that is achieved when projecting the high-dimensional data on the local low-dimensional manifolds. Since  (\ref{eq:eomc}-\ref{eq:eomc4}) is an unsupervised learning problem, we can not apply most of the standard hyper-parameter selection routines from ML and AI, like cross-validation and Bayesian hyper-parameter tuning \cite{wu19}. Instead, one can use here the hyper-parameter selection from the unsupervised regularization methods in statistics and computational science, like the $L$-curve method \cite{hansen93}. 

Alternatively, in the following examples we will adopt a nonparametric information-theoretic perspective to model selection - aiming to find the hyper-parameter combinations that lead to the models combining simplicity (measured as a high lossy compression rate, computed as a ratio between the raw data complexity and EOMC model descriptor length), and quality (measured as the relative loss of compression). In contrast to common parametric measures from information theory (like Akaike and Bayesian Information Criteria) \cite{burnham13} that rely on validity of parametric assumptions like Gaussianity, this non-parametric model selection procedure based on lossy compression allows a robust identification of good hyper-parameter combinations across all of the benchmarks considered below. As will also be shown on application examples below, EOMC does not require a careful and precise adjustment, since the results remain robust in the broad ranges of hyper-parameters.

\section{Application examples}
\label{sec:results}
Next, we will demonstrate several applications of the EOMC algorithm introduced above, in comparison with the state-of-the-art methods, for  the synthetic noisy data examples with known low-dimensional manifold structures, for the data produced by the simplified Lorenz-96 model from fluid mechanics, as well as for the much more realistic modified Hasegawa-Wakatani (mHW) model of drift-wave turbulence in the edge of a tokamak plasma. 

\subsection{Synthetic examples}
\label{sec:synt}
\subsubsection{Example 1: nonstationary mixture of data switching between 3D ball surface and 3D torus surface in 10 dimensions with noise} 
\label{sec:synt_ex1}
\begin{figure}[h!]
        \advance\leftskip-1cm
        \includegraphics[clip,  width=1.1\textwidth]{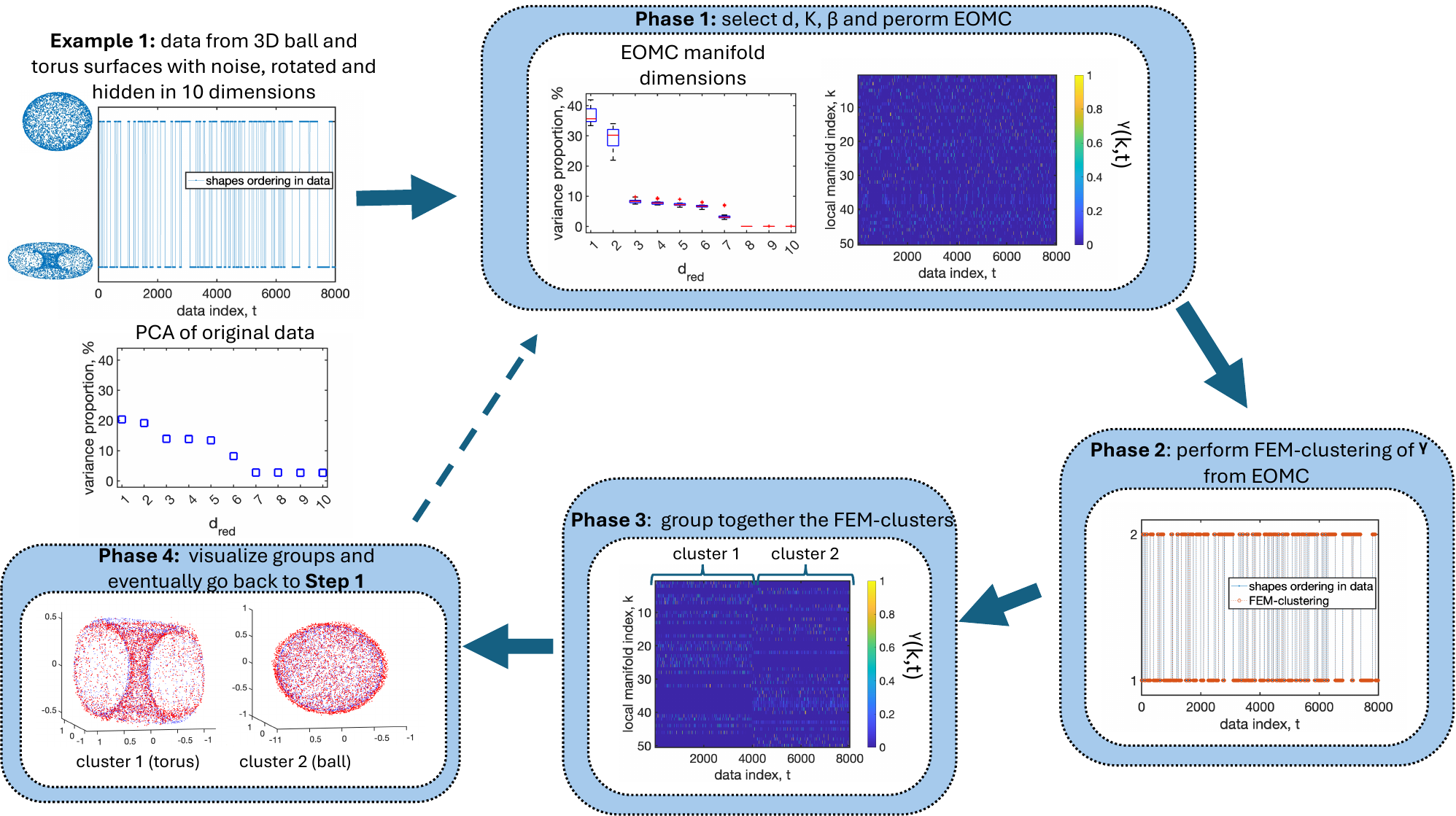}
 \caption{Graphic illustration of the five phases of the EOMC data analysis pipeline. Text description is provided in the Sec.~\ref{sec:synt_ex1}. Reconstructions of manifolds (red dots) are visualized applying the equation (\ref{eq:projectionl}).}
\end{figure}
\begin{figure}[h!]
 \centering
        \includegraphics[clip,  width=1.1\textwidth]{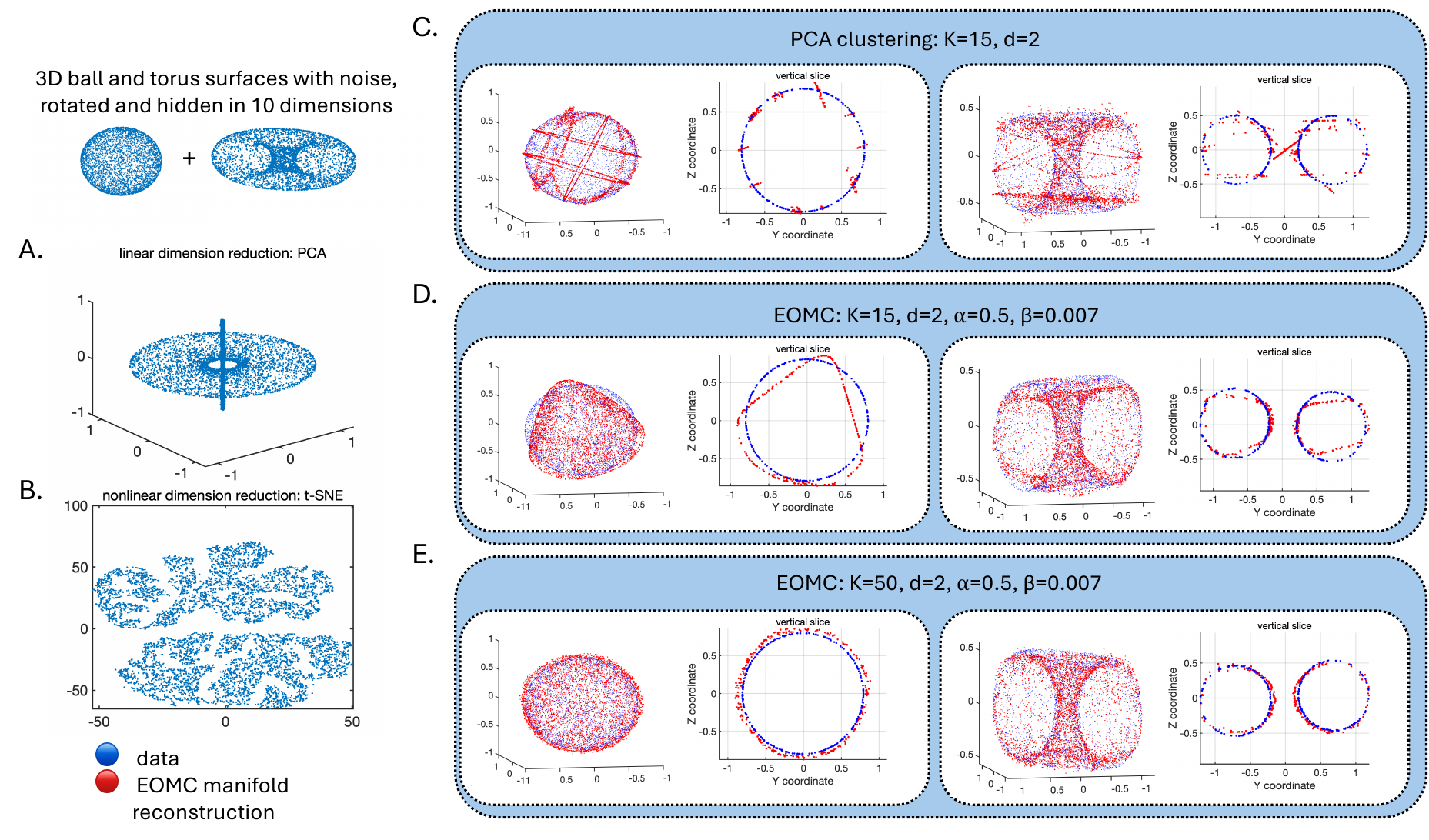}
 \caption{Analysis results for the nonstationary data from Example 1 (switching between two-dimensional ball and torus surface manifolds in ten dimensions with noise). Reconstructions of manifolds are visualized applying the equation (\ref{eq:projectionl}).}
\end{figure}
First, we will consider a non-stationary process (shown on the upper left of Fig.~1),  switching between the two non-linear manifolds of very different topology: a 3D torus surface and a 3D ball surface. To make the problem more challenging for manifold learning, generated 3D data matrix is further randomly rotated in 10 dimensions and subject to a 10-dimensional Gaussian noise.

We will use this first example also to illustrate the EOMC data analysis pipeline (graphically illustrated in Fig.~1), that will be also used in the following examples:
\begin{itemize}
\item {\bf Phase 1}: select the hyper-parameters $d,K, \alpha, \beta$ (and $\beta_0$, if simultaneous learning of data reliability $\gamma_0$ with (\ref{eq:eomc_g0}-\ref{eq:eomc_g06})  is incorporated in EOMC\footnote{ Here, and in the next example described in Sec.~\ref{sec:synt_ex2}, we perform the analysis setting $\gamma_0\equiv1/T$ and $\beta_0=0$. An example of EOMC with simultaneous learning of the entropy-optimal data reliability function $\gamma_0$ will be provided in Sec.~\ref{sec:synt_ex3}}), as described in Sec.~\ref{sec:hyper}. Next, perform the EOMC analysis according to the Theorem 1 above. Note that the application of the standard linear PCA (left panel in Fig.~1) does not reveal any significant and clear  spectral gap in the eigenvalues of the covariance, and does not indicate a presence of any low-dimensional manifold. In contrast, applying EOMC with arbitrarily selected $K=15$ indicates a clear spectral gap after the second eigenvalues of the localized EOMC covariance matrices $\mathrm{Cov}_k\left(X,\gamma,\mu_k^*\right)$. This clearly indicates that the data contains non-linear manifolds with local dimensionality $d=2$. 
\item {\bf Phase 2}: take the matrix of internal coordinates $\gamma$ produced in the  {\bf Phase 1}, and either use it directly, to reconstruct the low-dimensional representation with (\ref{eq:projectionl}) (i.e., going straight to {\bf Phase 4}), or subject $\gamma$ to further clustering, with any machine learning method that allows clustering of discrete probability distribution series, for example, with the Finite Element Method clustering (FEM-clustering) \cite{metzner12}, that deploys the cross-entropy as a distance metric for clustering the probability measures. As can be seen from the  {\bf Phase 2} illustration in the Fig.~1, this immediately uncovers the original switching process that was used in the data generation. Alternatively, as explained in the Sec.~\ref{sec:visual} below, one can omit the FEM-clustering and directly visualise the low-dimensional representation of the $\gamma$ with the standard tools of nonlinear t-SNE visualisation.  
\item {\bf Phase 3}: if FEM-clustering was deployed in  {\bf Phase 2}, in this step one brings together the manifold-projected data instances belonging to the same FEM-clusters, where projection is performed with the formula (\ref{eq:projectionl}).
\item {\bf Phase 4}:  visualise and inspect the obtained manifold reconstructions. When necessary, update the hyper-parameters and return to the  {\bf Phase 1}.
\end{itemize}         
Despite of the hyper-parameter adjustment, PCA and t-SNE fail to recover the two low-dimensional manifolds from these nonstationary data (see Fig.~2A and 2B). Fig.~2C illustrates further effects of hyper-parameter selection described in Sec.~\ref{sec:hyper}: setting both $\alpha=0$ and $\beta=0$ changes  (\ref{eq:eomc}-\ref{eq:eomc4}) to its special case, i.e., to PCA-clustering \cite{horenko06,horenko08,metzner12}. It makes visible the central problem of PCA-clustering, discussed above and induced by the Lemmas 1 and 2:   the kernel of the manifold projection operator is non-empty when $d<D$ (Lemma 1), and the errors orthogonal to the manifold are not visible to the method. Therefore, the approximation in this case  is rather cutting through the manifold than approximating it. Setting $\alpha$ and $\beta$ to some non-zero values mitigates this problem (see Fig.~2D). Then, increasing the number $K$ of local manifolds from 15 (which was the initial guess) to 50 results in almost perfect reconstruction of both nonlinear manifolds (see Fig.~2D). Please note that obtaining this result did not require a tedious hyper-parameter tuning, results shown in Fig.~2C were obtained from only two repetitions of the EOMC pipeline described in the Fig.~1.    
 \subsubsection{Example 2: nonstationary mixture of data switching between 1D peace sign contour (planar) and 1D  prism contour (in 3D), embedded and rotated in 100 dimensions with noise}  
\label{sec:synt_ex2}
\begin{figure}[h!]
 \centering
        \includegraphics[clip,  width=1.1\textwidth]{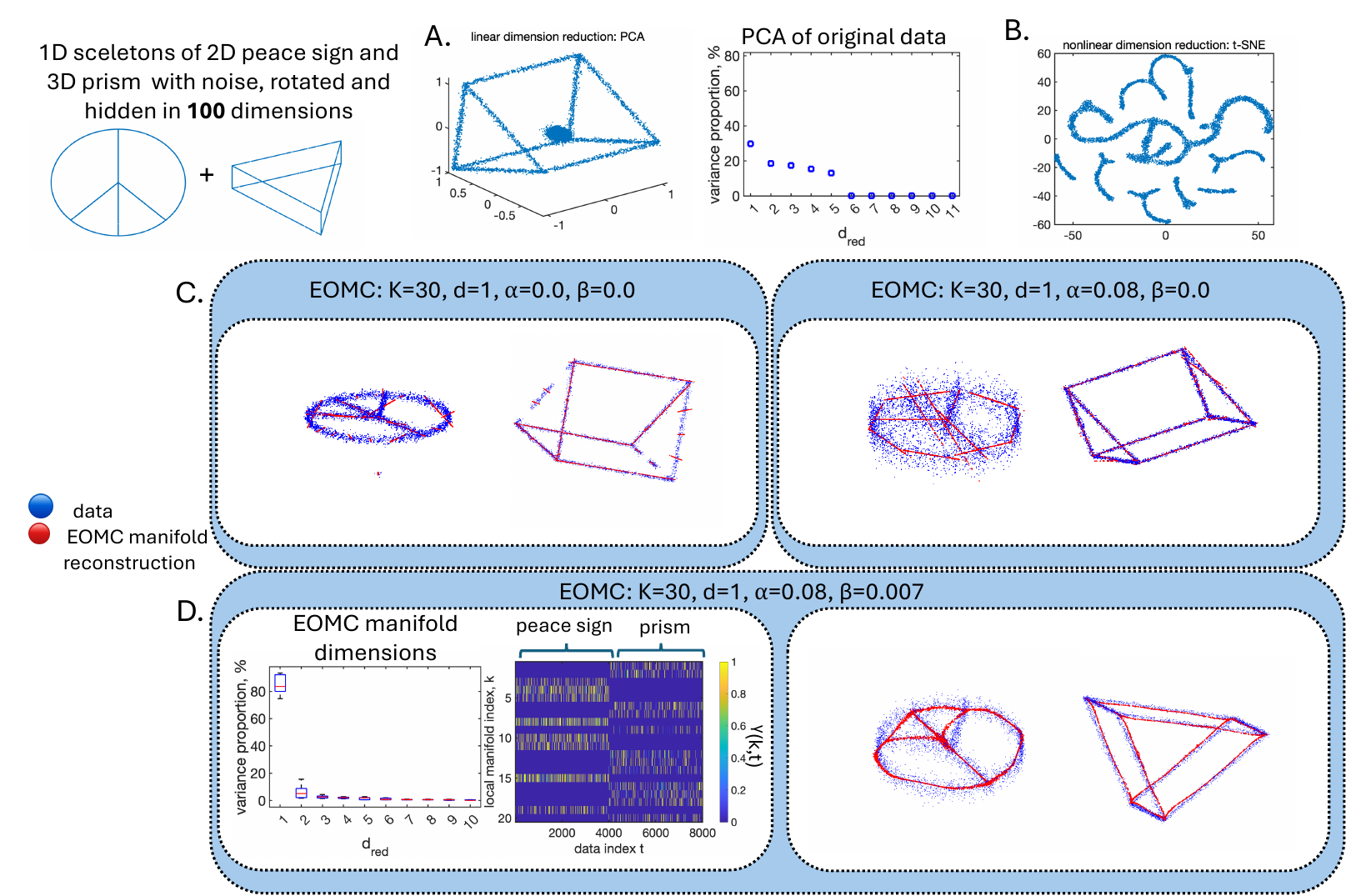}
 \caption{Analysis results for the nonstationary data from Example 2 (switching between one-dimensional peace sign and prism contours manifolds in hundred dimensions with noise). Reconstructions of manifolds (red dots) are visualized applying the equation (\ref{eq:projectionl}).}
\end{figure}
Next, we will consider  a non-stationary process  switching between the two non-linear one-dimensional manifolds: a 1D contour of a 3D prism and a 1D contour of the 2D peace sign, containing both piece-wise linear and nonlinear parts. To make the problem even more challenging for manifold learning (then in the previous example from Sec.~\ref{sec:synt_ex1}), generated 3D data matrix is further randomly rotated in 100 dimensions and subject to a 100-dimensional Gaussian noise.

As in the previous example, despite of the hyper-parameter tuning, both linear PCA and t-SNE fail to find these low dimensional manifolds hidden in 100 dimensions of noisy data (see Fig.~3A and 3B).  Instead, t-SNE finds a lot of clusters (see Fig.~3B), that are not present in the data, representing a clear artefact of t-SNE.  Fig.~3C illustrates again the problem of PCA-clustering, and Figs.~3D and 3E show the effect of the entropic regularization term in (\ref{eq:eomc}-\ref{eq:eomc4}): setting $\alpha>0$ and $\beta=0$ results in "edgy" piece-wise linear approximation of the nonlinear manifold fragments (Fig.~3D), whereas setting  $\alpha>0$ and $\beta>0$ allows obtaining much better and more smooth interpolations. As in the example 1 above, obtaining the result from the Fig.~3E  did not require a tedious hyper-parameter adjustment, requiring to go through the EOMC pipeline described in the Fig.~1 only once.

\subsubsection{Example 3: co-inference of input reliability $\gamma_0(t)=\bP\left[X(:,t)\in\Omega\right]$ in EOMC learning} 
\label{sec:synt_ex3}
Next, we illustrate the application of the modified EOMC (with a simultaneous learning the input data reliability $\gamma_0$), as described in Sec.~ \ref{sec:gamma_0} and eqs. (\ref{eq:eomc_g0}-\ref{eq:gamma0}). We generate a noisy 2D data (red crosses in Fig.~4A), that are normally distributed around a smooth 1D non-linear manifold following a smooth letter "S" (black line in Fig.~4A). As in the previous synthetic examples, to make the manifold learning more challenging, we add an additional 10-dimensional Gaussian noise and randomly rotate the training data in 10 dimensions. As discussed in Sec.~ \ref{sec:gamma_0}, the EOMC formulation in (\ref{eq:eomc}-\ref{eq:eomc4}) implicitly assumes that all of the training data points have the same $1/T$ contribution to the loss function - i.e., according to the uniform probability distribution prior. However, this assumption can be seriously violated when considering the data points not used in training and distributed uniformly inside of the whole 10-dimensional hypercube - and far away from the training data (blue dots in Fig.~4).  As can be seen from Fig.~4B, applying the projection formula (\ref{eq:projectionl}) from the Theorem 1 results in artefact projections that are far away of the true manifold (red circular dots in Fig.~4B). 

In contrast, co-inference of data reliability function  $\gamma_0(t)=\bP\left[X(:,t)\in\Omega\right]$ in the modified EOMC from Sec.~ \ref{sec:gamma_0} and eqs. (\ref{eq:eomc_g0}-\ref{eq:gamma0}) produces a very reliable reconstruction of the true manifold: the intensity of red for circular dots in Fig.~4C is proportional to the inferred values of data reliability $\gamma_0(t)$. The unreliable points far away from the training data and from the EOMC-approximated manifold automatically get values of $\gamma_0(t)$ close to zero. As in the previous two synthetic examples, achieving this result did not require any particular hyper-parameter tuning: it was obtained for an ad hoc choice of $K=6, d=1,  \alpha=5, \beta=0.3, \beta_0=4$. This result remains remarkably-robust, and does not change notably when changing these hyper-parameter values in quite broad ranges.    
\begin{figure}[h!]
 \centering
        \includegraphics[clip,  width=1.0\textwidth]{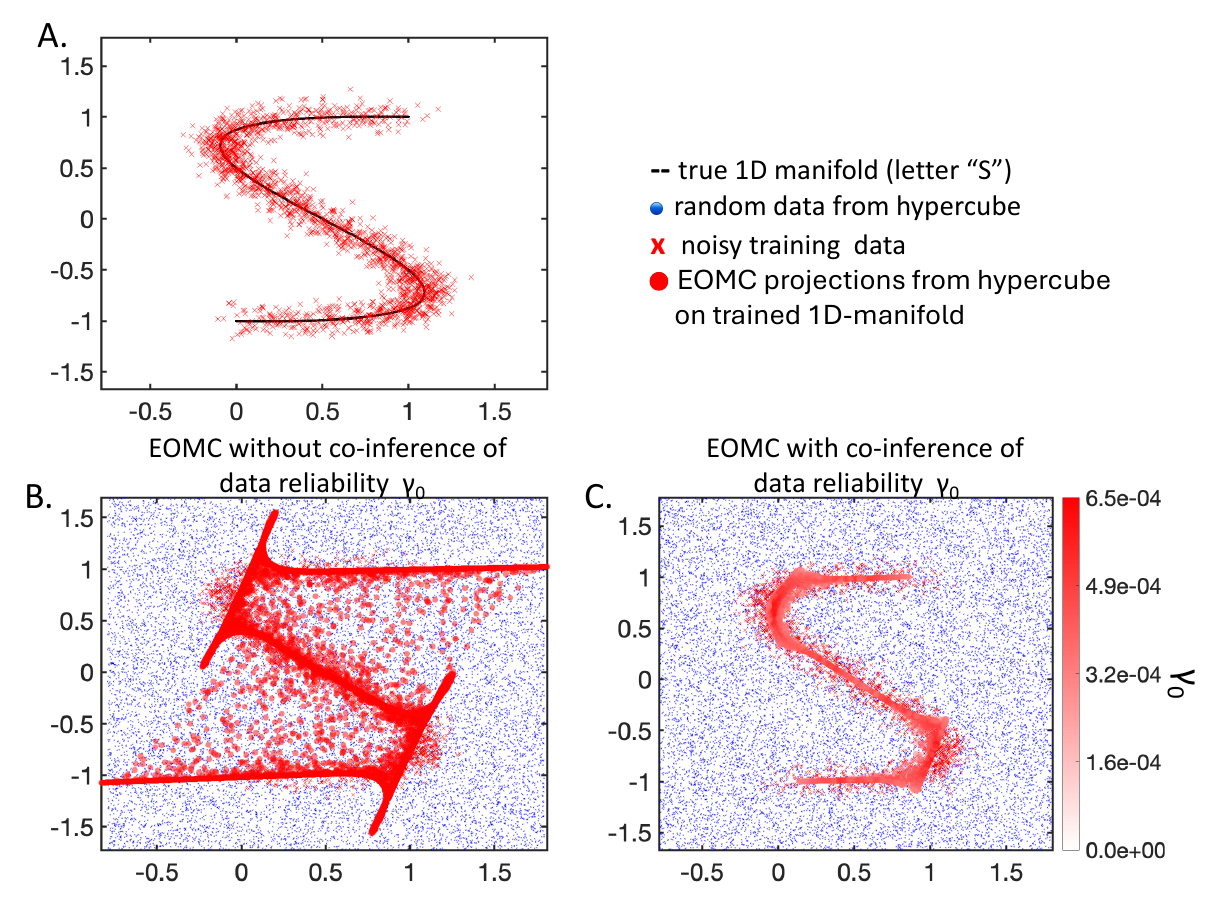}
 \caption{Illustration of the modified EOMC learning from Sec.~ \ref{sec:gamma_0} and eqs. (\ref{eq:eomc_g0}-\ref{eq:gamma0}): co-inference of data reliability function $\gamma_0$.}
\end{figure}

 \subsection{Analysis of data from Lorenz-63 in chaotic, strongly-chaotic and very strongly-chaotic regimes}
 \label{sec:lorenz}
\paragraph{Model description} The Lorenz-96 (L96) model was introduced by Edward Lorenz in a 1996 paper (published later in 2005) as a simplified, yet sophisticated, "toy model" of the Earth's atmosphere for studying the fundamental issues of predictability and chaotic dynamics in spatially extended systems \cite{Lorenz1996,Lorenz2005}. It mimics aspects of the mid-latitude atmosphere's non-linear dynamics, such as advection, dissipation, and external forcing, within a computationally cheap, periodic one-dimensional domain (a latitude circle). The L96 model is widely used today as a benchmark problem for data assimilation techniques, ensemble forecasting methods, and studies on the general nature of spatiotemporal chaos \cite{Lorenz1998,anderson2001,houtekamer2005,bocquet2020}.

The L96 Type 1 model consists of a system of $N$ coupled ordinary differential equations (ODEs), describing the time evolution of a single scalar atmospheric quantity $X_j$ at $N$ equally spaced grid points around a latitude circle:
\begin{equation}
\frac{dX_j}{dt} = (X_{j+1} - X_{j-2})X_{j-1} - X_j + F \quad \text{for } j = 1, \dots, N
\label{eq:L96}
\end{equation}
Periodic boundary conditions are assumed, such that indices are taken modulo $N$ (i.e., $X_{j+N} = X_j$ and $X_{j-N} = X_j$).
Variables and terms in (\ref{eq:L96}) have the following meaning:
\begin{itemize}
    \item $X_j$: The value of the atmospheric quantity (e.g., temperature, vorticity) at the $j$-th grid point.
    \item $N$: The total number of grid points in the system (system size). Common values in literature are $N=40$.
    \item $t$: Time.
    \item $F$: A positive, constant external forcing parameter that drives the system.
    \item $(X_{j+1} - X_{j-2})X_{j-1}$: The non-linear advection term, which conserves energy in the absence of forcing and damping.
    \item $-X_j$: A linear damping (dissipation) term.
\end{itemize}


Behaviour of the L96 model changes significantly with the forcing parameter $F$. For small values of $F$ (e.g., $F < 1$), the system exhibits periodic or steady-state dynamics. As $F$ increases, the system undergoes bifurcations and transitions into chaotic regimes. A commonly studied value is $F=8$, which produces robust chaotic behaviour used frequently as a standard benchmark in predictability studies. For regimes where $F \geq 7$ (which includes $F=9$ and $F=12$ investigated below), the system is considered to be in a strong or fully turbulent chaotic state \cite{Lorenz2005}. 

\paragraph{Application of EOMC to L96 output data} 
\begin{figure}[h!]
 \centering
        \includegraphics[clip,  width=1.\textwidth]{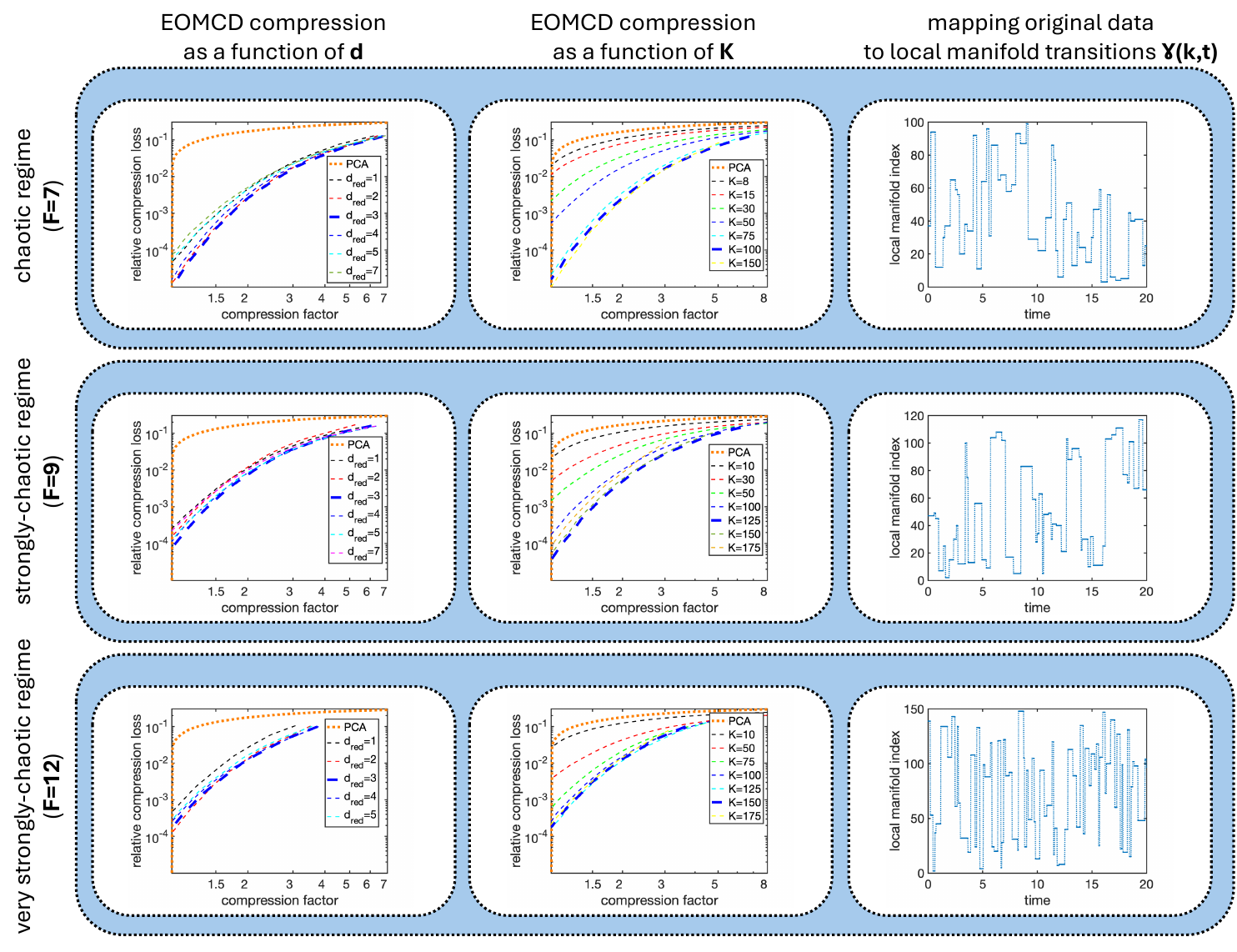}
 \caption{EOMC analysis results for the data from Lorenz-96 model with the external forcings $F=7,10,12$ (in rows) for the dependence between compression factor and loss as functions of reduced manifold dimensionality $d$ (first column) and number $K$ of local manifolds (second column), as well
  as the identified trajectories of EOMC internal coordinates $\gamma$ as functions of time (third column). Compression factor is computed as the ratio of the size of the original data $X$ to the number of scalar real values required to store the EOMC parameters $\left\{\gamma^*,\mu^*_1,\mathcal{T}^*_1,\dots,\mu^*_K,\mathcal{T}^*_K\right\}$, that  are sufficient to compute the reconstruction (\ref{eq:pca-clustering-reconst_full}) of X from its projection on the manifold, for different values of $K$ . Relative compression error is the relative l1-norm error of approximating the full data $X$ with its manifold  reconstruction (\ref{eq:pca-clustering-reconst_full}), for different values of $K$.   }
\end{figure}
\begin{figure}[h!]
 \centering
        \includegraphics[clip,  width=1.1\textwidth]{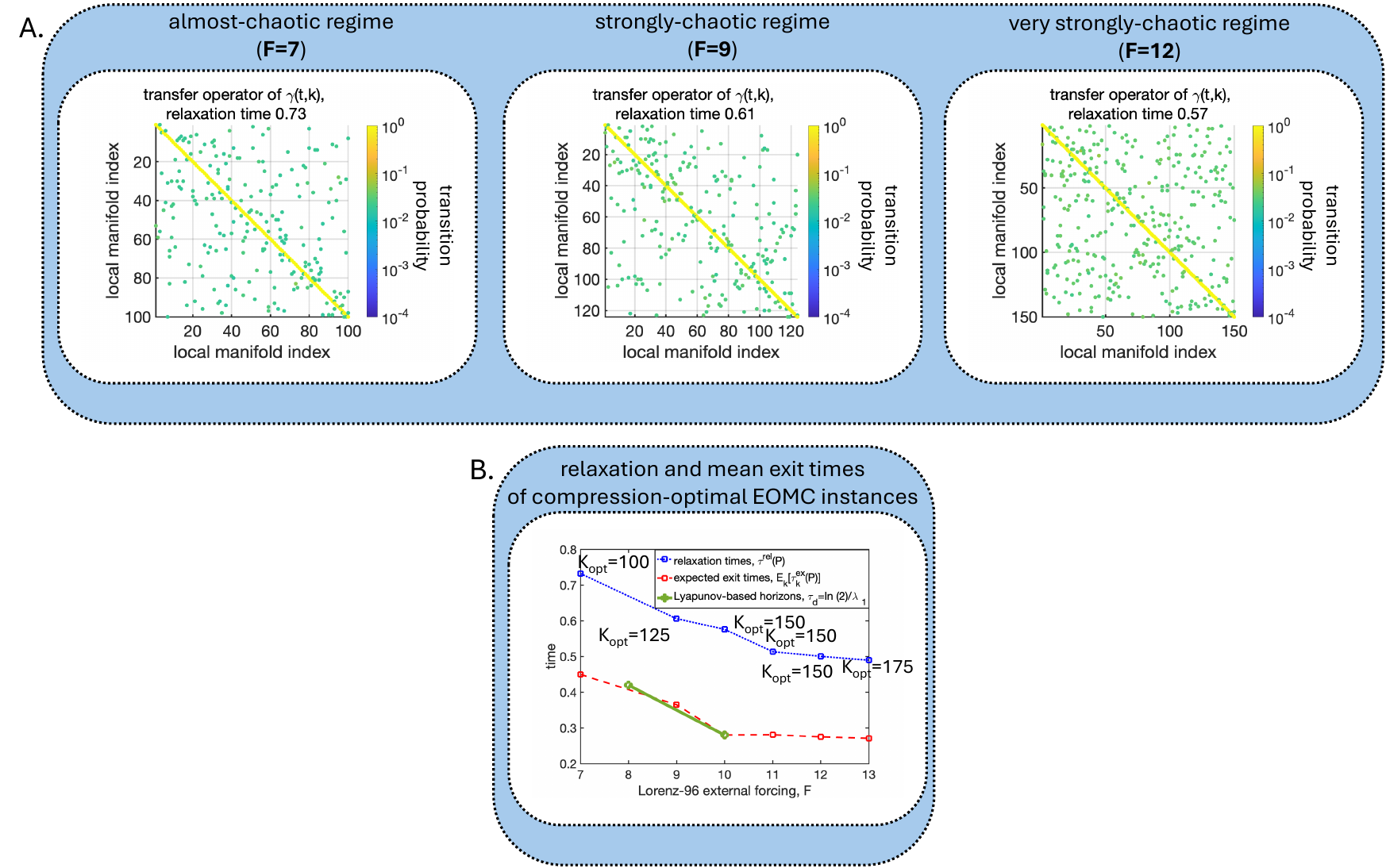}
 \caption{Panel A: heat maps of the transfer operator matrices inferred from the metastable EOMC $\gamma$ time series (partially shown in the third column of Fig.~4). Panel B: mean Markovian exit times and mean relaxation times for the Markovian transfer operators inferred for $\gamma$ that resulted from the 'EOMC analysis in a range of forcings $F$ covering the chaotic and strongly-chaotic regimes of the L96-model. }
\end{figure}
In the following, we will use the common literature setting for $N=40$, and generate long time series $X\in\mathcal{R}^{40\times30000}$ of L96 for the three forcing regimes $F=7,9,12$,  with $T=30000$ and time step $\tau=0.02$, covering the total period of 600  time units. After rescaling according to \cite{Lorenz1996,Lorenz1998}, this corresponds to around 3'000 Earth atmospheric days. Assuming ergodicity of the L96-model, these data should be enough to cover the realistic period of the atmosphere measurements.  In each of the forcing regimes we train EOMC in a broad range of hyper-parameter settings, applying the nonparametric lossy compression to find the optimal hyper-parameter values, as described above in the Sec.~2.4. As can be seen from the first and the second columns of Fig.~5, for a broad range of compression ranges, the minimal compression loss is achieved for $d_{opt}=3$ in all of the forcing regimes\footnote{Interestingly, literature mentioning optimal local dimensionality for L96-model is quite limited, and indicates that the local attractor dimension is very difficult to determine from data, and they see a broad range of values depending on how an approximation to it is constructed \cite{egusphere-2024-3915}.} , and for $K_{opt}$ (thick dashed blue lines in the first and the second columns of Fig.~5) going from $100$, over $125$ and to $150$ when $F$ goes from $7$, over $9$, and to $12$. Please note from the first and the second columns of Fig.~5 that at the same levels of compression factors as PCA, EOMC allows achieving almost two orders of magnitude smaller relative compression losses.  

We inspect the time series of intrinsic EONC coordinates $\gamma$ that were computed  with these optimal hyper-parameter choices. $\gamma$ encode the probabilities of points to belong to different local manifolds $\left\{\mu_k,\mathcal{T}_k\right\}$, $k=1,\dots,K$ at different time series instances. As can be seen from the  Fig.~5 and Fig.~6A, $\gamma$ exhibits very persistent dynamics in time - with probability to stay on the local manifold being almost two orders of magnitude larger than a probability to switch to another manifold - indicating relatively-long stays in each of the  local linear manifolds (with $d_{opt}=3$). Please note that this \emph{metastability} of $\gamma$ can not be an artefact of the EOMC data analysis: as can be seen from (\ref{eq:eomc}-\ref{eq:eomc4}), the EOMC loss function does not contain any terms that would enforce the persistence or metastability on $\gamma$. As a matter of fact, the value of $\mathcal{L}^{\textrm{EOMC}}$ in (\ref{eq:eomc}-\ref{eq:eomc4}) is invariant with respect to any permutation of the columns of data matrix $X$. This means that the observed  \emph{metastability} of $\gamma$ can only be an imprint of the underlying L96 dynamics - that appears to be best described by a metastable process switching between low-dimensional (with $d=3$) locally-linear manifolds.   Next, for each of the obtained $\gamma$, we compute the Markovian transition operators $P$ that describe the time evolution of $\gamma(:,t)$, by means of the exact law of the total probability, i.e., $\gamma(:,t+\tau)=P\gamma(:,t)$, where $P_{i,j}=\bP\left[\gamma(j,t+\tau)=1|\gamma(i,t)=1\right]$ \cite{schutte13}. Elements $P_{i,j}$ of the transfer operator $P$ contain the probabilities of transitions from state $j$ to state $i$ in a single time step $\tau$. As demonstrated by the  Fig.~6A, operators $P$  are characterized by high probabilities of staying in the states (large diagonal entries) and low probabilities of transitions to other local manifold  states (very small off-diagonal entries). Next, for each of the three transfer operators computed for each of the three L96 forcing regimes, we compute the relaxation times $\tau^{rel}(P)=\frac{\tau}{1-|\lambda_2|}$, where $\tau$ is the time step, and $\lambda_2$ is the second largest (in absolute value) eigenvalue of $P$ \cite{schutte13}. Relaxation times $\tau^{rel}(P)$ in Markov processes measure the predictability horizons: they quantify the time it takes for the Markov processes $P$ to forget the initial condition, and to converge towards the invariant density measure \cite{schutte03,schutte10,schutte13}. As can be seen from the Fig.~6A,  $\tau^{rel}(P)$ gradually reduces from  $0.73$, over $0.61$ and to $0.57$ when $F$ goes from $7$, over $9$, and to $12$. 

Next we investigate the behaviour of  $\tau^{rel}(P)$ and $\bE_{k}\left[\tau_{k}^{ex}(P)\right]$ (where $\tau^{ex}_{k}(P)=\frac{\tau}{1-P_{k,k}}$  is a Markovian mean exit time from state $k$, and $\bE_{k}$ is the mathematical expectation over all $k=1,\dots,K$) on a more dense grid of L96 forcings $F$ between $F=7$ and $F=13$ (see Fig.~6B). Mean exit times  $\tau^{ex}_{k}(P)$ quantify an average time that Markov process spends in a state $k$ before leaving it. In fluid mechanics and geosciences, predictability horizons are usually measured with the error doubling times \(\tau _{d}=\frac{\ln (2)}{\lambda _{1}}\), computed from a positive leading Lyapunov exponent $\lambda_1$. For typical atmospheric parameters simulated by the L96 model (e.g., $N=40, F=8$), the error doubling times reported in the literature are short, roughly corresponding to the short-term forecast limits observed in real-world weather prediction models (around 2-2.5 days in atmospheric terms, corresponding to 0.42 time units of L96 with $N=40, F=8$) \cite{Lorenz1996,Lorenz1998}. Larger forcings $F$ mean more chaotic behaviour and much shorter doubling times, e.g., for  $N=40, F=10$ doubling time and prediction horizon is 0.28 time units \cite{Olascoaga2010}. As can be seen from the red curve in Fig.~6B, average Markovian mean exit times  $\bE_{k}\left[\tau_{k}^{ex}(P)\right]$ closely match these doubling times from positive Lyapunov exponents in the literature - indicating the predictability horizons quantified with Lyapunov exponents as limits for the average times it takes for the system to leave the current low-dimensional manifold and to go somewhere else. Transfer operator description revealed by the EOMC analysis in Figs.~5 and 6 goes beyond this limit: besides allowing to measure the average time $\tau_{k}^{ex}(P)$ the dynamics spends in the local manifold $k$, transfer operator provides the transition probabilities $P_{j,k}$ to all of the other local manifold states $j\neq k$, where the dynamics can go to after leaving $k$. As revealed by the relaxation times curve $\tau^{rel}(P)$ (see the blue dotted curve in the Fig.~6B), this transfer operator description roughly doubles the prediction horizon for the system, as compared to the state of the art Lyapunov exponent description.      

\subsection{Analysis of data from modified Hasegawa-Wakatani (mHW) model of drift-wave turbulence in the edge of a tokamak plasma.}
 \label{sec:mHW}
 One-dimensional L96-model is considered by many to provide only a very limited, and simplified description of turbulent atmospheric dynamics - and not a model of a "real" turbulence behaviour. To check if the findings from previous Section about metastability and predictability horizons of L96 are induced by this oversimplification - or if they are also reflecting intrinsic properties of "real" turbulent systems, next we will consider an application of EOMC to the output of a more realistic model from Magnetohydrodynamics (MHD).  
We will take the Hasegawa-Wakatani model - a seminal two-field fluid description model of drift-wave turbulence in the edge of a tokamak plasma \cite{hasegawa83,wakatani84,gottwald04}. It reduces the complex MHD equations to the evolution of density fluctuations (\(n\)) and electrostatic potential (\(\phi \)) fields. The modified Hasegawa-Wakatani (mHW) model describes the evolution of electrostatic potential $\phi$ and density fluctuations $n$ in a two-dimensional slab geometry. In the following application example, we will deploy the mHW model version introduced by Numata et al. \cite{numata2007}, involving the resistive coupling term acting only on non-zonal fluctuations.

\paragraph{Governing Equations} 
Let the zonal average of a field $f$ be defined as $\langle f \rangle = \frac{1}{L_y} \int f \, dy$, and the non-zonal fluctuation as $\tilde{f} = f - \langle f \rangle$. The mHW equations are:
\begin{align}
    \frac{\partial \zeta}{\partial t} + \{\phi, \zeta\} &= \alpha (\tilde{\phi} - \tilde{n}) - \mu \nabla^4 \zeta, \label{eqn:vorticity} \\
    \frac{\partial n}{\partial t} + \{\phi, n\} + \kappa \frac{\partial \phi}{\partial y} &= \alpha (\tilde{\phi} - \tilde{n}) - \mu \nabla^4 n, \label{eqn:density}
\end{align}
where $\zeta = \nabla^2 \phi$ is the ion vorticity and $\{a, b\} = \partial_x a \partial_y b - \partial_y a \partial_x b$ is the Poisson bracket representing $\mathbf{E} \times \mathbf{B}$ advection.
Variables and parameters of mHW model equations \cite{numata2007} have the following meaning:
\begin{itemize}
\item    $\phi$: electrostatic potential;
\item     $n$: electron density fluctuations;
\item     $\zeta$: ion vorticity ($\nabla^2 \phi$);
\item     $\alpha$: adiabaticity parameter (resistive coupling);
\item     $\kappa$: background density gradient scale length ($-\partial_x \ln n_0$);
\item     $\mu$: dissipation/viscosity coefficient.
\end{itemize}
\paragraph{Data generation}
To generate the time series for the EOMC analysis, we use the MATLAB code by Jean-Christophe Nave and Denis St-Onge, available at \url{https://github.com/DenSto/HWE_solver}.  We use the same settings for all of the mWH model parameters as in this code, with the only change being a slightly reduced grid size ($64\times 64$ grid points). We generate a time series on the interval $\left[600,3000\right]$ (skipping the outputs between $t=0$ and $t=600$, when the model  "burns-in" and the transient states die-out), with constant time intervals $\delta t=0.3125$. Finally, we create a data matrix $X$ for EOMC, containing the time series instances for electron density fluctuation $n$ and ion vorticity $\zeta$ fields (see Fig.~7A). 
\begin{figure}[h!]
 \centering
        \includegraphics[clip,  width=1.1\textwidth]{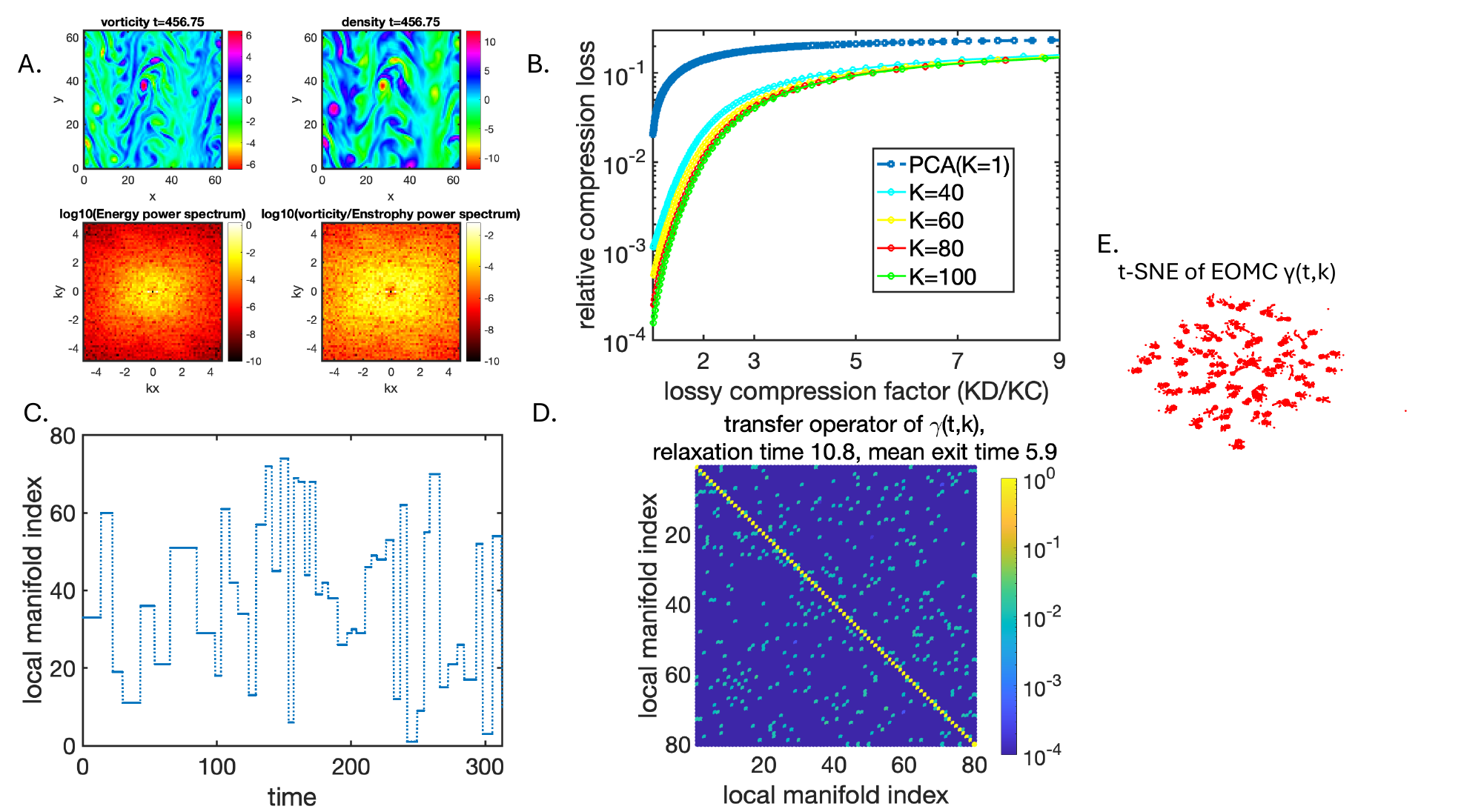}
 \caption{Application of EOMC to the time series output of modified 2D Hasegawa-Wakatani model (mHW) \cite{hasegawa83,wakatani84,gottwald04}. For details of the description see Sec.~\ref{sec:mHW}.}
\end{figure}
\paragraph{Application of EOMC to mHW output data} 

We follow the same analysis procedure as deployed for  the previous L96-example in Sec.~\ref{sec:lorenz}: first, we determine optimal values for the reduced manifold dimension $d$ (that appears to be $d=4$), and $K$ (that appears to be $K=80$, see Fig.~7B). The optimal values for the other hyper-parameters appear to be $\alpha=0.1, \beta=10^{-3}, \beta_0=1.0$ - and as in the previous examples, obtained results appear to remain robust in broad ranges of these hyper-parameter values. As can be seen when comparing Fig.~7B with the middle column panels of Fig.~5, similarly to L96, for the same level of lossy data compression, application of EOMC to mHW data allows achieving over around 15-fold smaller relative lossy compression errors then PCA.   Fig.~7C shows that the optimal $\gamma(k,t)$ appears to be a discrete, metastable and persistent  transition process - staying long times in the same local low-dimensional manifolds before switching to the next ones. Finally - and very similarly to the previous L96-example (see Fig.~5 and 6) - Markovian transition probability matrix $P$ obtained from $\gamma(k,t)$ appears to be very metastable, with mean exit times of  5.9 and the predictability horizon (measured as the mean relaxation time of $P$) being almost twice  as long - around 10.8 time units.

\section{Visualising EOMC internal coordinates $\gamma$ with t-SNE} 
\label{sec:visual}
\begin{figure}[h!]
 \centering
       \includegraphics[clip,  width=1.\textwidth]{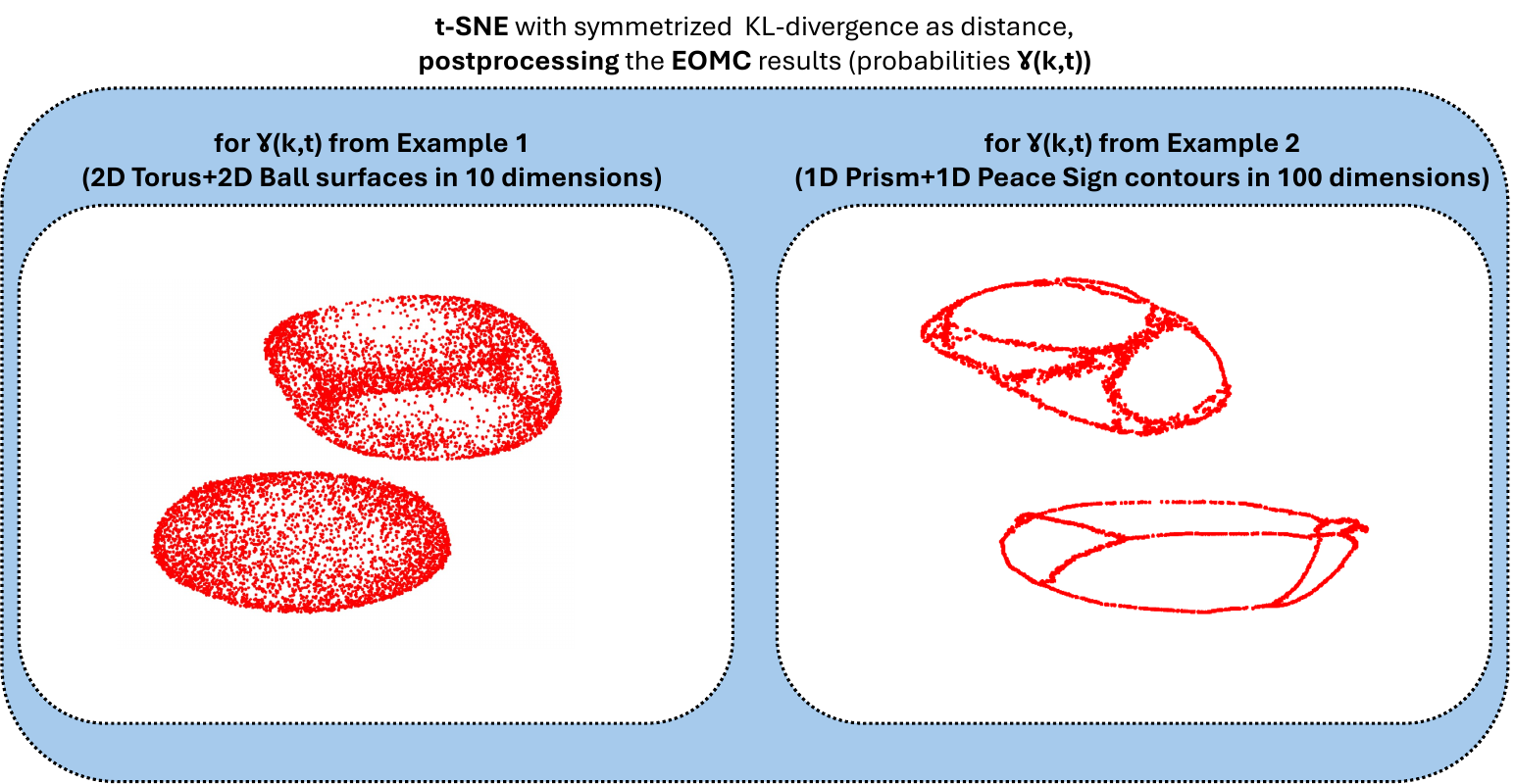}
        \includegraphics[clip,  width=0.8\textwidth]{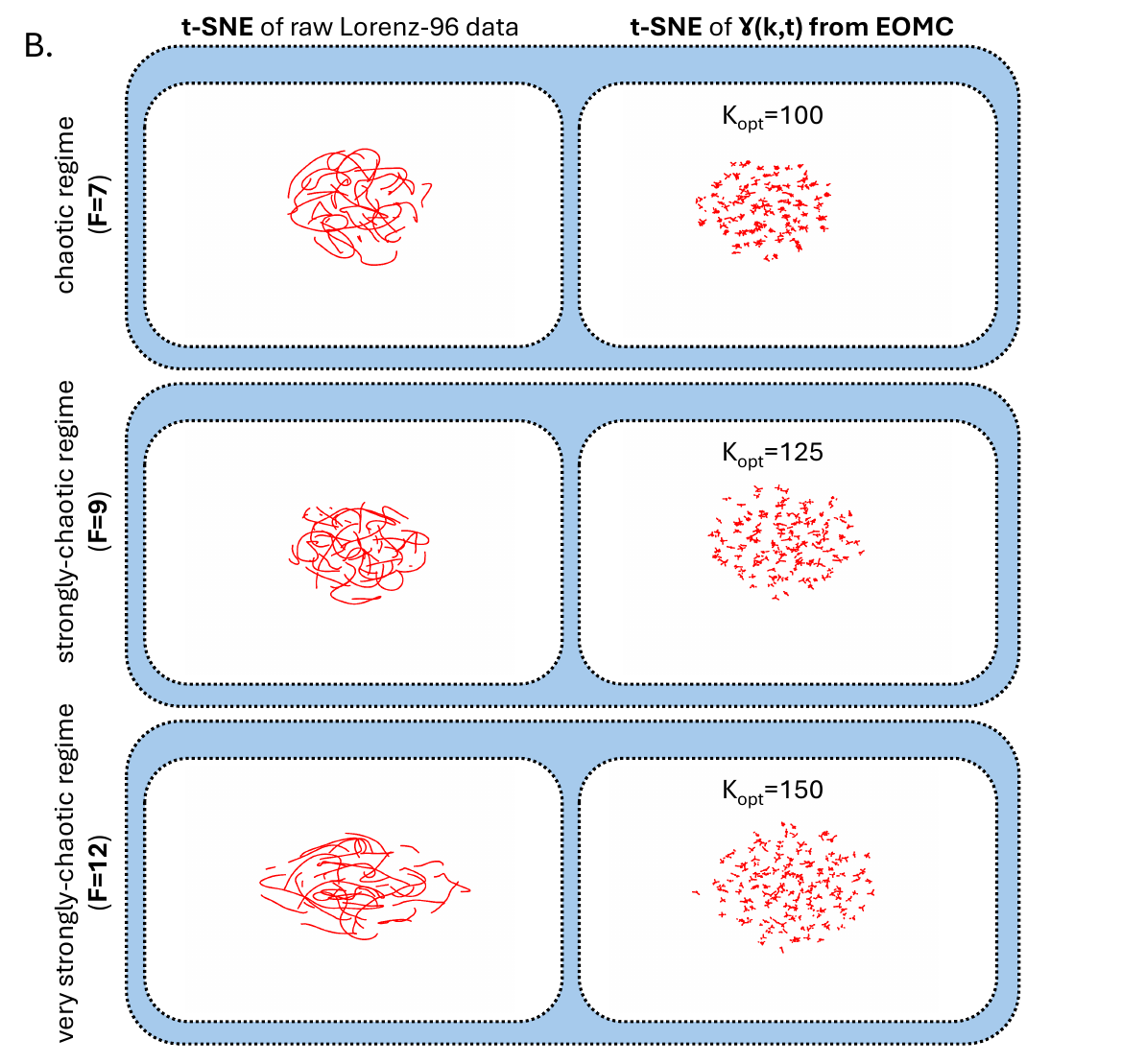}
 \caption{Nonlinear t-SNE for visualisation of the EOMC internal coordinates $\gamma$ from  (\ref{eq:eomc}-\ref{eq:eomc4}). Panel A: visualising $\gamma$ from the Example 1 in Sec.~\ref{sec:synt_ex1} (left), and  $\gamma$ from the Example 2 in Sec.~\ref{sec:synt_ex2}. Panel B: t-SNE visualisation of raw data from the L96 model described in Sec.~\ref{sec:lorenz} (left column), and for EOMC $\gamma$ (right column).} 
\end{figure}
As mentioned above, state-of-the-art nonlinear methods like t-SNE and UMAP rely on tuning of a multitude of hyper-parameters, both method- and numerics-specific. However, in examples 1 and 2 from Sections \ref{sec:synt_ex1} and \ref{sec:synt_ex1}, t-SNE failed to recover the low-dimensional manifolds from these nonstationary and nonlinear data (see Figs.~2B and 3B). Instead, in both of the cases it produced multiple clusters that were not present in the generated data.

Alternatively, we can deploy t-SNE to directly visualise in two or three dimensions the structures like the EOMC interpolation coefficients $\gamma$. For this, in t-SNE we can not use the common Euclidean metrics, but need to deploy a distance measure that is conform with probability measures, for example, the cross-entropy or the symmetrized Kullback-Leibler divergence.  As shown in the Fig.~8A, without a particular hyper-parameter tuning (by just setting the perplexity parameter of t-SNE somewhere in the range of values between 400 and 1'200) allows fully recovering all of the original low-dimensional manifolds used in the data generation, although with some distortion. This result is a bit surprising, since it does not require taking the EOMC manifold projections from the formula (\ref{eq:projectionl}) - and indicates that the topological structure of the manifolds is already contained in the EOMC probabilities  $\gamma$.

Applying t-SNE to the raw data from the Lorenz-96 model in Sec.~\ref{sec:lorenz} reveals no pronounced clusters - although they appear to be very prominent when applying the EOMC (see Figs.~5 and 6). In contrast, without any particular hyper-parameter tuning, t-SNE with the symmetrized KL-distance reveals a very pronounced cluster structure in the $\gamma$ from EOMC. In accordance with the other results obtained in the Sec.~\ref{sec:lorenz}, t-SNE reveals that the number of low-dimensional manifold clusters gradually grows with the growth of the L96 extern forcing $F$ - but the topology of these manifolds looks very similar,  differing only in their position and orientation.        

Applying t-SNE with a  symmetrized Kullback-Leibler divergence as distance to the EOMC interpolation coefficients $\gamma$ from mHW example analysis in Sec.~\ref{sec:mHW} results in a very similar picture - with multiple clearly distinct clusters (see Fig.~7E) of very similar topology,  differing only in their position and orientation.

\section{Discussion}
\label{sec:discuss}
Data-driven manifold learning and dimensionality reduction methods are constituting one of the central pillars of data analysis in many areas of science. For example, in neurosciences and bioinformatics, nonlinear methods like t-SNE and UMAP are very widely used to investigate and to visualise the high-dimensional data structures, as well as to detect clusters. However, examples provided above illustrate that t-SNE - one of the most popular tools, with over 28'000 citations according to Google Scholar - can create clusters that are not present in the data (see Figs.~2B and 3B), as well as to not detect clusters when they are actually present (see the left panels of Fig.~8B).  Another problems of such methods include polynomial, or, in the best cases, $\mathcal{O}\left(T\log\left(T\right)\right)$ cost scaling with the size $T$ of data statistics - as well as necessity to tune multiple model-specific and numerics-specific hyper-parameters, and with no consensus on the optimal procedure needed to select them. Last but not least, state-of-the-art manifold learning and nonlinear dimensionality reduction methods do not provide straightforwardly-computable measures of input data reliability - reducing robustness of these approaches and making them vulnerable to the adversarial attacks. 

As demonstrated in the Sec. \ref{sec:eomc} and proven in the Lemmas 3, 4, and in the Theorem 1, Entropy Optimal Manifold Clustering (EOMC) allows mitigating these problems, resulting in computational cost scaling of $\mathcal{O}\left(T\right)$, with an explicit rule  (\ref{eq:projectionl}) to project the new data points on the manifold with cost scaling of $\mathcal{O}\left(D\right)$ in the leading order, providing a very robust learning of very nonlinear manifolds from noisy and nonstationary data, not requiring complicated hyper-parameter adjustment (as shown on the synthetic examples the Sec.~\ref{sec:synt_ex1} and \ref{sec:synt_ex2}).  Moreover, as was shown in the Sec.~\ref{sec:gamma_0}, deploying the explicit analytical solution (\ref{eq:gamma0}), one can easily modify EOMC without an increase of leading order computational cost - and allow simultaneous learning of the data  reliability measure $\gamma_0$. As  demonstrated in Sec.~\ref{sec:synt_ex3} and Fig.~4, this modification of EOMC allows a much more robust and artefact-free functioning of the method (compare Fig.~4B to Fig.~4C).     

 
 Before we discuss the results obtained for Lorenz-96 from fluid mechanics when applying the EOMC method proposed in this paper (see Sec.~\ref{sec:lorenz}), we will briefly recapitulate the current knowledge regarding topology and predictability bounds in the chaotic and strongly-chaotic regimes of the Lorenz-96 model:

\begin{itemize}
    \item \textbf{High-Dimensional Chaos:} As forcing $F$ increases, the system exhibits extensive spatiotemporal chaos. The fractal dimension of the attractor grows, indicating a large number of active chaotic degrees of freedom, although the number of relevant dimensions may saturate in the strong driving limit \cite{Patil2001, Olascoaga2010}.
    \item \textbf{Finite Predictability Limit:} Like the real atmosphere it mimics, the L96 model in these regimes shows sensitive dependence on initial conditions, leading to a finite predictability horizon \cite{Lorenz2005}. Errors grow exponentially over time, characterized by a positive leading Lyapunov exponent $\lambda_1$.
    \item \textbf{Error Doubling Time:} Studies often quantify predictability horizons for a system in terms of the error doubling time \(\tau _{d}=\frac{\ln (2)}{\lambda _{1}}\), computed from a positive leading Lyapunov exponent $\lambda_1$. For typical atmospheric parameters simulated by the model (e.g., $N=40, F=8$), the error doubling time is short, roughly corresponding to the short-term forecast limits observed in real-world weather prediction models (around 2-2.5 days in atmospheric terms, corresponding to 0.42 time units of L96 with $N=40, F=8$) \cite{Lorenz1996,Lorenz1998}. Larger forcings $F$ mean more chaotic behaviour and much shorter doubling times, e.g., for  $N=40, F=10$ doubling time and prediction horizon is 0.28 time units \cite{Olascoaga2010}
    \item \textbf{Predictability Bounds:} While the theoretical \textit{intrinsic} predictability of the atmosphere might be around two weeks, the L96 model often reproduces \textit{practical} predictability limits (e.g., 4-5 days of useful forecasts) depending on the resolution $N$ and the forcing $F$ used in the specific experiment. The strong $F \geq 7$ regimes are characterized by very rapid error growth, making accurate long-term forecasting impossible without perfect models and initial conditions \cite{Lorenz2005}.
\end{itemize}

From the current perspective, predictability in weather systems or inhomogeneous turbulence / geophysical fluids more generally, is typically limited by the emergence of coherent structures. Predictability in the observed weather systems is then considered to be a local property, with all the complexity that it brings.  For the high-dimensional dynamical systems,  Lyapunov exponents and local Kaplan-Yorke dimensions are among the popular measures of local dimensionality and predictability \cite{okane25}.

The results obtained in the Sec.~\ref{sec:lorenz} tell a somewhat different story. EOMC analysis reveals that the internal coordinates $\gamma$ in all of the considered forcing regimes, exhibit very persistent and \emph{metastable} dynamics in time, indicating relatively long stays in each of the  local linear manifolds (with relatively low dimensionality $d=3$). Similar metastable behaviour for a simplified barotropic quasi-geostrophic ODE model of atmospheric dynamics was previously demonstrated applying Hidden Markov Model (HMM) analysis from machine learning, and reported in the work of A. Majda et al \cite{majda06}. However, HMMs impose a Markovian structure and Markov-property on the data during analysis procedure (i.e., the property that every next observation in time series is conditionally-dependent on the previous one). The value of Markovian functional used in the HMM training is not invariant with respect to permutations of the data, so also the order of the data in time series matters for the HMM analysis outcome.  In other words, the metastability of a simplified barotropic quasi-geostrophic ODE model reported in \cite{majda06} could be alternatively explained as an artefact of HMM data analysis.

In contrast, as was analysed above in Sec.~\ref{sec:lorenz},  \emph{metastability} of $\gamma$ can not be an artefact of the EOMC data analysis: as can be seen from (\ref{eq:eomc}-\ref{eq:eomc4}), the EOMC loss function does not contain any terms that would enforce the persistence, metastability or Markov-property on $\gamma$. As a matter of fact, the value of $\mathcal{L}^{\textrm{EOMC}}$ in (\ref{eq:eomc}-\ref{eq:eomc4}) is invariant with respect to any permutation of the columns of data matrix $X$. This means that the observed  \emph{metastability} of $\gamma$ for Lorenz-96 in chaotic and very chaotic regimes can only be an imprint of the underlying L96 dynamics - that appears to be best described by a metastable process switching between low-dimensional (with $d=3$) locally-linear manifolds. Analysis from Sec.~\ref{sec:lorenz} revealed that the main effect of increasing forcing was in gradually-increasing the total number $K$ of these local low-dimensional manifolds - and in slowly decreasing mean relaxation and mean exit times, that were measured from the transfer operators inferred from EOMC variables $\gamma$  (see Fig.~6). We shown that average Markovian mean exit times  $\bE_{k}\left[\tau_{k}^{ex}(P)\right]$ closely match the doubling times from positive Lyapunov exponents in the literature (see Fig.~6B) - indicating the predictability horizons quantified with Lyapunov exponents as limits for the average times it takes for the system to leave the current low-dimensional manifold and to go somewhere else. And, it was shown that the transfer operator description revealed by the EOMC analysis in Figs.~5 and 6 allows going beyond this limit: besides allowing to measure the average time $\tau_{k}^{ex}(P)$ the dynamics spends in the local manifold $k$, transfer operator provides the transition probabilities $P_{j,k}$ to all of the other local manifold states $j\neq k$, where the dynamics can go to after leaving $k$. For Markov processes, relaxation times define the predictability horizon of the system, quantifying the time that the system requires to forget its initial condition \cite{schutte13}. As revealed by the relaxation times curve $\tau^{rel}(P)$ (see the blue dotted curve in the Fig.~6B), this transfer operator description roughly doubles the prediction horizon for the system, as compared to the state of the art Lyapunov exponent description.  

However, one-dimensional L96-model provides only a very limited, and a very simplified description of turbulent atmospheric dynamics. To check if the findings  about metastability and predictability horizons of L96 are induced solely by this oversimplification - or if they are also reflecting intrinsic properties of "real" turbulent systems, next we considered an application of EOMC to the output of a more realistic model from Magnetohydrodynamics (MHD):  we took the modified Hasegawa-Wakatani model (mHW) - a seminal two-field fluid description model of drift-wave turbulence in the edge of a tokamak plasma \cite{hasegawa83,wakatani84,gottwald04}. As can be seen from the results provided in the Sec.~\ref{sec:mHW} and from comparisons of Figs.~5 and 6 to a Fig.~7, results for this much more advanced mHW model tell basically the same story as the L96 results: EOMC reveals that the dynamics can be described by a metastable process infrequently switching between the different local low-dimensional manifolds, with the predictability horizon given by the mean relaxation times, almost doubling the prediction horizons measured with the mean exit times and Lyapunov exponents. In other words, metastability  is not an artefact of L96-simplification, but is an essential characteristics of much more complex turbulent models like mHW, that do not rely on L96-simplifications. 

These findings open very exciting possibilities for applying various very advanced tools from transfer operator research to the areas of fluid mechanics and geosciences. Potentially-useful approaches include methods like adaptive transfer operator sampling, milestoning, Markovian transition pathways theory and algorithms, and many others \cite{metzner06,pham2006,schutte23}.

\bmhead{Availability of code}  Code can be shared upon a reasonable request. 

\bmhead{Acknowledgement} The author would like to thank Davide Bassetti and Tim Prokosch (both RPTU Kaiserslautern-Landau), Lukas Pospisil (VTU Ostrava), Michael Groom and Terry O'Kane (both CSIRO), Georg Gottwald (U Sydney)  and the other participants of the Oberwolfach Mini-Workshop  "Mathematics of Entropic AI in the Natural Sciences" in 2024, discussions with whom provided a lot of motivation for picking-up this work. Christoph Sch\"utte and Rupert Klein (both FU Berlin)  pioneered transfer operator theory and PCA-clustering that provided foundations for this work. Results on Lorenz-96 long-term predictability in chaotic regime due to the metastable manifold switching were build upon the hypothesis originally formulated by Andrew J. Majda (NYU, deceased in 2021) during the private discussion in 2010 at IPAM.
This work was funded by the EU Horizons project $AI4LUNGS$ (Grant Agreement No. 101080756).

\bibliography{MAD}

@article{egusphere-2024-3915,
	author = {Bonte, M. and Vannitsem, S.},
	doi = {10.5194/egusphere-2024-3915},
	journal = {EGUsphere},
	pages = {1--39},
	title = {Finite-size local dimension as a tool for extracting geometrical properties of attractors of dynamical systems},
	url = {https://egusphere.copernicus.org/preprints/2024/egusphere-2024-3915/},
	volume = {2024},
	year = {2024}}

@article{okane25,
author = {Axelsen, Andrew R. and O’Kane, Terence J. and Quinn, Courtney R. and Bassom, Andrew P.},
title = {Hyperbolicity and Southern Hemisphere Persistent Synoptic Events},
journal = {Journal of Advances in Modeling Earth Systems},
volume = {17},
number = {4},
pages = {e2024MS004834},
doi = {https://doi.org/10.1029/2024MS004834},
url = {https://agupubs.onlinelibrary.wiley.com/doi/abs/10.1029/2024MS004834},
eprint = {https://agupubs.onlinelibrary.wiley.com/doi/pdf/10.1029/2024MS004834},
note = {e2024MS004834 2024MS004834},
year = {2025}
}

@article{majda06,
	author = {Andrew J. Majda and Christian L. Franzke and Alexander Fischer and Daniel T. Crommelin},
	doi = {10.1073/pnas.0602641103},
	eprint = {https://www.pnas.org/doi/pdf/10.1073/pnas.0602641103},
	journal = {Proceedings of the National Academy of Sciences},
	number = {22},
	pages = {8309-8314},
	title = {Distinct metastable atmospheric regimes despite nearly Gaussian statistics: A paradigm model},
	url = {https://www.pnas.org/doi/abs/10.1073/pnas.0602641103},
	volume = {103},
	year = {2006}}

@article{wakatani84,
  title={A collisional drift wave description of plasma edge turbulence},
  author={Wakatani, M. and Hasegawa, A.},
  journal={Physics of Fluids},
  volume={27},
  number={3},
  pages={611--618},
  year={1984}
}

@article{gottwald04,
  title={Arakawa-like schemes for the Hasegawa-Wakatani equations},
  author={Gottwald, G. A. and Grimshaw, R.},
  journal={Journal of Computational Physics},
  volume={197},
  number={1},
  pages={210--231},
  year={2004}
}

@article{hasegawa83,
  title={Plasma Edge Turbulence},
  author={Hasegawa, A. and Wakatani, M.},
  journal={Physical Review Letters},
  volume={50},
  number={9},
  pages={682--686},
  year={1983}
}

@article{numata2007,
  title = {A modified Hasegawa-Wakatani model for plasma turbulence and zonal flows},
  author = {Numata, R. and Ball, R. and Dewar, R. L.},
  journal = {Physics of Plasmas},
  volume = {14},
  number = {10},
  pages = {102312},
  year = {2007},
  publisher = {AIP Publishing}
}

@article{metzner06,
  title={Illustration of transition path theory on a collection of simple examples},
  author={Metzner, Philipp and Sch{\"u}tte, Christof and Vanden-Eijnden, Eric},
  journal={The Journal of chemical physics},
  volume={125},
  number={8},
  year={2006},
  publisher={AIP Publishing}
}

@article{pham2006,
  title={Robust fusion of irregularly sampled data using adaptive normalized convolution},
  author={Pham, Tuan Q and Van Vliet, Lucas J and Schutte, Klamer},
  journal={EURASIP Journal on Advances in Signal Processing},
  volume={2006},
  number={1},
  pages={083268},
  year={2006},
  publisher={Springer}
}

@article{schutte23,
  title={Overcoming the timescale barrier in molecular dynamics: Transfer operators, variational principles and machine learning},
  author={Sch{\"u}tte, Christof and Klus, Stefan and Hartmann, Carsten},
  journal={Acta Numerica},
  volume={32},
  pages={517--673},
  year={2023},
  publisher={Cambridge University Press}
}

@article{anderson2001,
  author = {Anderson, Jeffrey L.},
  year = {2001},
  title = {An Ensemble Adjustment Kalman Filter for Data Assimilation},
  journal = {Monthly Weather Review},
  volume = {129},
  number = {12},
  pages = {2884--2903},
  doi = {10.1175/1520-0493(2001)129<2884:AEAKFF>2.0.CO;2}
}

@article{houtekamer2005,
  author = {Houtekamer, P. L. and Mitchell, H. L.},
  year = {2005},
  title = {A Sequential Ensemble Kalman Filter for Atmospheric Data Assimilation},
  journal = {Monthly Weather Review},
  volume = {133},
  number = {5},
  pages = {1238--1250},
  doi = {10.1175/MWR2955.1}
}

@article{bocquet2020,
  author = {Bocquet, Marc and Brajard, Julien and Carrassi, Alberto and Bertino, Laurent},
  year = {2020},
  title = {Bayesian inference of chaotic dynamics by merging data assimilation, machine learning and expectation-maximization},
  journal = {Foundations of Data Science},
  volume = {2},
  number = {1},
  pages = {55--80},
  doi = {10.3934/fods.2020004}
}

@article{schutte03,
  title={Biomolecular conformations can be identified as metastable sets of molecular dynamics},
  author={Sch{\"u}tte, Christof and Huisinga, Wilhelm},
  journal={Handbook of numerical analysis},
  volume={10},
  pages={699--744},
  year={2003},
  publisher={Elsevier}
}

@inproceedings{schutte10,
  title={On Markov state models for metastable processes},
  author={Djurdjevac, Natasa and Sarich, Marco and Sch{\"u}tte, Christof},
  booktitle={Proceedings of the International Congress of Mathematicians 2010 (ICM 2010) (In 4 Volumes) Vol. I: Plenary Lectures and Ceremonies Vols. II--IV: Invited Lectures},
  pages={3105--3131},
  year={2010},
  organization={World Scientific}
}

@book{schutte13,
  title={Metastability and Markov state models in molecular dynamics.},
  author={Sch{\"u}tte, Christof and Sarich, Marco},
  volume={24},
  year={2013},
  publisher={American Mathematical Soc.},
address={New York, NY, USA},}

@article{Lorenz1998,
  author = {Lorenz, Edward N. and Emanuel, Kerry A.},
  year = {1998},
  title = {Optimal Sites for Supplementary Weather Observations: Experiments with a Small Model},
  journal = {Journal of the Atmospheric Sciences},
  volume = {55},
  number = {3},
  pages = {399--414},
  doi = {10.1175/1520-0469(1998)055<0399:OSFSWO>2.0.CO;2}
}

@article{Lorenz2005,
  author = {Lorenz, Edward N.},
  year = {2005},
  title = {Designing Chaotic Models},
  journal = {Journal of the Atmospheric Sciences},
  volume = {62},
  number = {5},
  pages = {1574--1587},
  doi = {10.1175/JAS3430.1}
}

@article{Lorenz1996,
  author = {Lorenz, Edward N.},
  year = {1996},
  title = {Predictability: a problem partly solved},
  booktitle = {Proceedings of the Seminar on Predictability},
  publisher = {ECMWF},
  address = {Reading, UK},
  pages = {1--18},
  }

@article{Patil2001,
  author = {Patil, D. J. and Frenkel, M. and Kermode, R. I.},
  year = {2001},
  title = {Chaos in the Lorenz 96 model: A thorough numerical study},
  journal = {International Journal of Chaos Theory and Applications},
  volume = {6},
  pages = {5--26}
}

@article{Olascoaga2010,
  author = {Olascoaga, Maria Jos{\'{e}} and Balachandar, S.},
  year = {2010},
  title = {Extensive chaos in the Lorenz-96 model},
  journal = {Chaos: An Interdisciplinary Journal of Nonlinear Science},
  volume = {20},
  number = {4},
  pages = {043105},
  doi = {10.1063/1.3496397},}

@book{burnham13,
  title={Model selection and inference: a practical information-theoretic approach},
  author={Burnham, K. and Anderson, D.},
  year={2013},
  publisher={Springer Science and Business Media},
address={New York, NY, USA},}

@article{hansen93,
  title={The use of the L-curve in the regularization of discrete ill-posed problems},
  author={Hansen, Per Christian and O’Leary, Dianne Prost},
  journal={SIAM journal on scientific computing},
  volume={14},
  number={6},
  pages={1487--1503},
  year={1993},
  publisher={SIAM}
}

@article{wu19,
  title={Hyperparameter optimization for machine learning models based on Bayesian optimization},
  author={Wu, Jia and Chen, Xiu-Yun and Zhang, Hao and Xiong, Li-Dong and Lei, Hang and Deng, Si-Hao},
  journal={Journal of Electronic Science and Technology},
  volume={17},
  number={1},
  pages={26--40},
  year={2019},
  publisher={Elsevier}
}

@article{bassetti25,
  title={An entropy-optimal path to humble AI},
  author={Bassetti, Davide and Posp{\'\i}{\v{s}}il, Luk{\'a}{\v{s}} and Groom, Michael and O'Kane, Terence J and Horenko, Illia},
  journal={arXiv preprint arXiv:2506.17940},
  year={2025}
}

@article{qp_np,
	author = {Sahni, Sartaj},
	doi = {10.1137/0203021},
	eprint = {https://doi.org/10.1137/0203021},
	journal = {SIAM Journal on Computing},
	number = {4},
	pages = {262-279},
	title = {Computationally Related Problems},
	url = {https://doi.org/10.1137/0203021},
	volume = {3},
	year = {1974}}

@article{Tenenbaum2000Global,
  author = {Tenenbaum, Joshua B and De Silva, Vin and Langford, John C},
  journal = {Science},
  title = {{A global geometric framework for nonlinear dimensionality reduction}},
  volume = {290},
  number = {5500},
  pages = {2319--2323},
  year = {2000},
  publisher = {{American Association for the Advancement of Science}},
  url = {science.sciencemag.org},
}

@article{Roweis2000Nonlinear,
  author = {Roweis, Sam T and Saul, Lawrence K},
  journal = {Science},
  title = {{Nonlinear dimensionality reduction by locally linear embedding}},
  volume = {290},
  number = {5500},
  pages = {2323--2326},
  year = {2000},
  publisher = {{American Association for the Advancement of Science}},
  url = {science.sciencemag.org},
}

@article{vanderMaaten2008Visualizing,
  author = {van der Maaten, Laurens and Hinton, Geoffrey},
  journal = {{Journal of Machine Learning Research (JMLR)}},
  title = {{Visualizing Data using t-SNE}},
  volume = {9},
  pages = {2579--2605},
  year = {2008},
  }

@article{umap_param,
  title={Parametric UMAP embeddings for representation and semisupervised learning},
  author={Sainburg, Tim and McInnes, Leland and Gentner, Timothy Q},
  journal={Neural Computation},
  volume={33},  number={11},
  pages={2881--2907},
  year={2021},
  publisher={MIT Press One Rogers Street, Cambridge, MA 02142-1209, USA journals-info~…}
}

@article{umap_interp,
  title={scGen predicts single-cell perturbation responses},
  author={Lotfollahi, Mohammad and Wolf, F Alexander and Theis, Fabian J},
  journal={Nature methods},
  volume={16},
  number={8},
  pages={715--721},
  year={2019},
  publisher={Nature Publishing Group US New York}
}

@article{umap_kernel,
  title={Sampling-enabled scalable manifold learning unveils the discriminative cluster structure of high-dimensional data},
  author={Peng, Dehua and Gui, Zhipeng and Wei, Wenzhang and Li, Fa and Gui, Jie and Wu, Huayi and Gong, Jianya},
  journal={Nature Machine Intelligence},
  pages={1--16},
  year={2025},
  publisher={Nature Publishing Group UK London}
}

@article{majda12,
	author = {Dimitrios Giannakis and Andrew J. Majda},
	doi = {10.1073/pnas.1118984109},
	eprint = {https://www.pnas.org/doi/pdf/10.1073/pnas.1118984109},
	journal = {Proceedings of the National Academy of Sciences},
	number = {7},
	pages = {2222-2227},
	title = {Nonlinear Laplacian spectral analysis for time series with intermittency and low-frequency variability},
	url = {https://www.pnas.org/doi/abs/10.1073/pnas.1118984109},
	volume = {109},
	year = {2012}}

@article{Becht2018Evaluating,
  author = {Becht, Etienne and McInnes, Leland and Healy, John and Dutertre, Charles-Antoine and Kwok, Ian W H and Ng, Lai Guan and Ginhoux, Florent and Newell, Evan W},
  journal = {{Nature Biotechnology}},
  title = {{Evaluating the manifold topology of single-cell data using UMAP}},
  volume = {37},
  number = {1},
  pages = {38--44},
  year = {2019},
  url = {www.nature.com}
}

@article{Ma2011Manifold,
  author = {Ma, Yin and Derksen, Harm},
  editor = {Ma, Yin and Derksen, Harm},
  title = {{Manifold Learning Theory and Applications}},
  year = {2011},
  publisher = {Taylor \& Francis},
  isbn = {978-1439871096}
}

@article{Sun2019Comparative,
  author = {Sun, Wen and Qu, Jiangda and Sun, Xiaoting and Fu, Kun and Meng, Deyu and Ngan, King},
  title = {{A Comparative Review of Manifold Learning Techniques for Hyperspectral Image Classification}},
  journal = {{Remote Sensing}},
  volume = {11},
  number = {6},
  pages = {681},
  year = {2019},
  publisher = {{MDPI}},
}

@article{McInnes2018,
  doi = {10.21105/joss.00861},
  url = {https://doi.org/10.21105/joss.00861},
  year = {2018},
  publisher = {The Open Journal},
  volume = {3},
  number = {29},
  pages = {861},
  author = {McInnes, Leland and Healy, John and Saul, Nathaniel and Gro{\ss}berger, Lukas},
  title = {{UMAP: Uniform Manifold Approximation and Projection}},
  journal = {{Journal of Open Source Software}}
}

@article{Healy2024UMAP,
  author = {Healy, John and McInnes, Leland},
  title = {{Uniform manifold approximation and projection}},
  journal = {{Nature Reviews Methods Primers}},
  volume = {4},
  number = {1},
  pages = {82},
  year = {2024},
  publisher = {{Nature Publishing Group}},
  url = {www.nature.com}
}

@article{vanderMaaten2014Accelerating,
  author = {van der Maaten, Laurens},
  journal = {{Journal of Machine Learning Research}},
  title = {{{Accelerating t-SNE using tree-based algorithms}}},
  volume = {15},
  number = {1},
  pages = {3221--3245},
  year = {2014},
}

@book{Jolliffe2002PCA,
  author = {Jolliffe, Ian T.},
  title = {Principal Component Analysis},
  edition = {2nd},
  year = {2002},
  publisher = {Springer},
  address = {New York, NY},
  series = {Springer Series in Statistics},
  isbn = {978-0387954424},
  doi = {10.1007/b98835}
}

@article{TabPFN25,
	author = {Hollmann, Noah and M{\"u}ller, Samuel and Purucker, Lennart and Krishnakumar, Arjun and K{\"o}rfer, Max and Hoo, Shi Bin and Schirrmeister, Robin Tibor and Hutter, Frank},
	date = {2025/01/01},
	doi = {10.1038/s41586-024-08328-6},
	id = {Hollmann2025},
	isbn = {1476-4687},
	journal = {Nature},
	number = {8045},
	pages = {319--326},
	title = {Accurate predictions on small data with a tabular foundation model},
	url = {https://doi.org/10.1038/s41586-024-08328-6},
	volume = {637},
	year = {2025}}

@inproceedings{horenko06,
	address = {Berlin, Heidelberg},
	author = {Horenko, Illia and Schmidt-Ehrenberg, Johannes and Sch{\"u}tte, Christof},
	booktitle = {Computational Life Sciences II},
	editor = {R. Berthold, Michael and Glen, Robert C. and Fischer, Ingrid},
	isbn = {978-3-540-45768-8},
	pages = {74--85},
	publisher = {Springer Berlin Heidelberg},
	title = {Set-Oriented Dimension Reduction: Localizing Principal Component Analysis Via Hidden Markov Models},
	year = {2006}}

@article{horenko08,
	author = {Horenko, Illia and Klein, Rupert and Dolaptchiev, Stamen and Sch\"{u}tte, Christof},
	doi = {10.1137/060670535},
	eprint = {https://doi.org/10.1137/060670535},
	journal = {Multiscale Modeling \& Simulation},
	number = {4},
	pages = {1125-1145},
	title = {Automated Generation of Reduced Stochastic Weather Models I: Simultaneous Dimension and Model Reduction for Time Series Analysis},
	url = {https://doi.org/10.1137/060670535},
	volume = {6},
	year = {2008}}

@article{metzner12,
  title={Analysis of persistent nonstationary time series and applications},
  author={Metzner, Philipp and Putzig, Lars and Horenko, Illia},
  journal={Communications in Applied Mathematics and Computational Science},
  volume={7},
  number={2},
  pages={175--229},
  year={2012},
  publisher={Mathematical Sciences Publishers}
}

@article{horenko_pnas_23,
author = {Illia Horenko  and Edoardo Vecchi  and Juraj Kardoš  and Andreas Wächter  and Olaf Schenk  and Terence J. O’Kane  and Patrick Gagliardini  and Susanne Gerber },
title = {On cheap entropy-sparsified regression learning},
journal = {Proceedings of the National Academy of Sciences},
volume = {120},
number = {1},
pages = {e2214972120},
year = {2023},
doi = {10.1073/pnas.2214972120},
URL = {https://www.pnas.org/doi/abs/10.1073/pnas.2214972120},
eprint = {https://www.pnas.org/doi/pdf/10.1073/pnas.2214972120},
}

@article{Horenko_2020,
	doi = {10.1162/neco\_a\_01296},
	year = 2020,
	month = {aug},
	publisher = {{MIT} Press - Journals},
	volume = {32},
	number = {8},
	pages = {1563--1579},
	author = {Illia Horenko},
	title = {On a Scalable Entropic Breaching of the Overfitting Barrier for Small Data Problems in Machine Learning},
	journal = {Neural Computation}
}

@article{espa_22,
    author = {Vecchi, Edoardo and Pospíšil, Lukáš and Albrecht, Steffen and O'Kane, Terence J. and Horenko, Illia},
    title = "{eSPA+: Scalable Entropy-Optimal Machine Learning Classification for Small Data Problems}",
    journal = {Neural Computation},
    volume = {34},
    number = {5},
    pages = {1220-1255},
    year = {2022},
    month = {04},
    issn = {0899-7667},
    doi = {10.1162/neco_a_01490},
    url = {https://doi.org/10.1162/neco\_a\_01490},
    eprint = {https://direct.mit.edu/neco/article-pdf/34/5/1220/2008663/neco\_a\_01490.pdf},
}

@article{horenko_pnas_22,
author = {Illia Horenko },
title = {Cheap robust learning of data anomalies with analytically solvable entropic outlier sparsification},
journal = {Proceedings of the National Academy of Sciences},
volume = {119},
number = {9},
pages = {e2119659119},
year = {2022},
doi = {10.1073/pnas.2119659119},
URL = {https://www.pnas.org/doi/abs/10.1073/pnas.2119659119},
eprint = {https://www.pnas.org/doi/pdf/10.1073/pnas.2119659119},
}

@article{horenko_17a,
	author = {Susanne Gerber and Illia Horenko},
	doi = {10.1073/pnas.1612619114},
	eprint = {https://www.pnas.org/doi/pdf/10.1073/pnas.1612619114},
	journal = {Proceedings of the National Academy of Sciences},
	number = {19},
	pages = {4863-4868},
	title = {Toward a direct and scalable identification of reduced models for categorical processes},
	url = {https://www.pnas.org/doi/abs/10.1073/pnas.1612619114},
	volume = {114},
	year = {2017}}

\end{document}